\documentclass[prd,preprintnumbers,aps,preprint,amsmath,amssymb,superscriptaddress,floatfix,nofootinbib,unsortedaddress]{revtex4}
\pdfoutput=1

\usepackage{amssymb, bm, amsmath, graphicx, physymb}
\usepackage[colorlinks=true,linktocpage=true,linkcolor=blue,citecolor=blue]{hyperref}
\usepackage[usenames,dvipsnames]{color}
\usepackage{slashed}
\usepackage[utf8]{inputenc}
\usepackage[T1]{fontenc}
\usepackage{enumerate}
\usepackage{amsfonts}
\usepackage{mathtools}
\usepackage{latexsym}
\usepackage{hyperref}
\usepackage{float}
\usepackage{epstopdf} % COMPILE EPS (AND PDF) FIGURES USING PDFLATEX!!!
\usepackage{bbm}
\usepackage{soul}
\usepackage[normalem]{ulem}
\usepackage{braket}
\usepackage{nccmath}
\usepackage{graphicx}
\usepackage{xcolor}

\usepackage[e]{esvect}

\def\cP{{\cal P}}

\newcommand{\secn}[1]{Section~1}
\newcommand{\appn}[1]{Appendix~1}

\long\def\comment#1{ }

\def\and{\quad\text{and}\quad}

\def\q{{\boldsymbol q}}
\def\0{{\boldsymbol 0}}
\def\1{{\boldsymbol 1}}
\def\p{{\boldsymbol p}}
\def\l{{\boldsymbol l}}

\def\y{{\boldsymbol y}}

\def\Y{{\boldsymbol Y}}
\def\r{{\boldsymbol r}}

\def\v{{\boldsymbol v}}

\def\0{{\boldsymbol 0}}

\def\max{{\rm max}}

\renewcommand\d{\delta}
\newcommand{\qv}{\q}
\newcommand{\pv}{\p}
\newcommand{\lv}{\l}

\newcommand{\tvec}{\boldsymbol}

\renewcommand{\part}{{\rm part}}
\newcommand{\be}{\begin{equation}}
\newcommand{\ee}{\end{equation}}
\newcommand{\bes}{\begin{subequations}}
\newcommand{\ees}{\end{subequations}}
\newcommand{\bea}{\begin{eqnarray}}
\newcommand{\eea}{\end{eqnarray}}

\newcommand{\pa}{\partial}

\newcommand{\nn}{\nonumber \\}
\newcommand{\na}{\nabla}
\newcommand{\beq}{\begin{equation}}
\newcommand{\eeq}{\end{equation}}

\makeatletter
\newsavebox{\@brx}
\newcommand{\llangle}[1][]{\savebox{\@brx}{\(\m@th{#1\langle}\)}%
  \mathopen{\copy\@brx\mkern2mu\kern-0.9\wd\@brx\usebox{\@brx}}}
\newcommand{\rrangle}[1][]{\savebox{\@brx}{\(\m@th{#1\rangle}\)}%
  \mathclose{\copy\@brx\mkern2mu\kern-0.9\wd\@brx\usebox{\@brx}}}
\makeatother

\begin{document}

\preprint{CERN-TH-2025-003}
\title{Non-local high-$p_t$ transport in anisotropic QCD matter }

\begin{abstract}
We perform a numerical study of non-local partonic transport in anisotropic QCD matter, relevant to the evolution of hard probes in the aftermath of high-energy nuclear scattering events.
The recently derived master equation, obtained from QFT considerations, differs from Boltzmann transport by incorporating a non-local elastic scattering kernel arising from density gradients.
After rewriting the master equation in a form suitable for numerical implementation and assuming a static density profile, we compare the non-local evolution to Boltzmann transport, demonstrating that the new interaction kernel is essential for accurately describing the azimuthal structure of the final-state momentum distribution. We further study the non-local partonic transport in the case of a matter profile 
governed by two-dimensional hydrodynamics, accounting for its flow and generalizing the
evolution equation. Our results demonstrate the necessity of going beyond classical transport at high-$p_t$ 
to accurately capture the structure of jets propagating through structured QCD matter. The master equation used in the numerical simulations can be seamlessly integrated into state-of-the-art transport codes.

\end{abstract}

\author{Jo\~{a}o Barata}
\email[Email: ]{joao.lourenco.henriques.barata@cern.ch}
\affiliation{CERN, Theoretical Physics Department, CH-1211 Geneva 23, Switzerland}

\author{Xiaojian Du}
\email[Email: ]{xiaojian.du@usc.es}
\affiliation{Instituto Galego de F{\'{i}}sica de Altas Enerx{\'{i}}as,  Universidade de Santiago de Compostela, Santiago de Compostela 15782, Galicia, Spain}

\author{Andrey V. Sadofyev}
\email[Email: ]{sadofyev@lip.pt}
\affiliation{LIP, Av. Prof. Gama Pinto, 2, P-1649-003 Lisboa, Portugal}

\maketitle

%\tableofcontents

\pagebreak

\section{Introduction}\label{sec:intro}

The collisions of ultrarelativistic heavy ions at the Large Hadron Collider (LHC) and Relativistic Heavy-Ion Collider (RHIC) lead to the formation of small droplets of the quark-gluon plasma (QGP), a quantum chromodynamic (QCD) state of matter consisting of free quarks and gluons, which populated our universe at its early stages. After being formed, these QGP droplets expand hydrodynamically,
driven by pressure gradients until the temperature is sufficiently low, at which point the hadronic remnants of the QGP free stream to the experimental detectors, see e.g.~\cite{Busza:2018rrf} for a recent review. In parallel with the evolution of this QCD medium, high-momentum scattering events can occur during collisions, producing energetic (outgoing) partons, which later generate collimated particle cascades, known as jets, resulting from the fragmentation of the initial parton. Since these branching processes occur while the medium is evolving, jets have been argued to be ideal probes of the QGP, offering a window to construct a spacetime picture of heavy-ion collisions~\cite{Apolinario:2024equ,Apolinario:2022vzg,Cao:2020wlm,Mehtar-Tani:2013pia}.

One of the key elements necessary to develop jet tomography of matter is the accurate description of medium-induced modifications to the jet structure. On the theory side, the major focus over the past decades has been on understanding how the successive interactions between the jet particles and the medium constituents result in the broadening of the jet spectrum and the emission of stimulated radiation. Focusing on the former effect and at leading order in the coupling, jet broadening results from the transverse momentum accumulated by the jet's partons due to interactions with the medium constituents, for recent discussions see e.g.~\cite{Mueller:2016xoc,Barata:2020rdn,Hauksson:2021okc,Sadofyev:2021ohn,Boguslavski:2023waw,Singh:2024pwr} and references therein. In the limit of a homogeneous and isotropic background, and for small momentum transfers between the jet and the medium--as detailed below--single parton evolution can be described in terms of a diffusive process controlled by a single transport coefficient: the jet quenching parameter $\hat q$. Focusing on the evolution of a single parton, the distribution $\cP(\p)$, which characterizes the accumulated transverse momentum $\p$ due to in-medium propagation, satisfies
\begin{align}\label{eq:diff_simple}
&\left(\pa_L-\frac{\hat{q}}{4}\pa_\p^2\right) \cP(\pv,L)=0\, ,
\end{align}
and $L$ refers to the propagation time along the parton trajectory.

In reality, the QGP produced in heavy-ion collisions is far from a uniform and isotropic state, as it is characterized by the presence of spatial gradients and collective flow~\cite{Busza:2018rrf}. As such, a complete theoretical description of the single parton evolution in a flowing anisotropic medium at leading order in QCD cannot be fully captured by Eq.~\eqref{eq:diff_simple} or its straightforward generalizations. Therefore, the theoretical description of jet evolution in matter requires a more careful treatment\footnote{In fact, there were earlier attempts to account for the transverse medium evolution in a phenomenologically motivated way, see e.g. \cite{Gyulassy:2001kr,Armesto:2004pt,Armesto:2004vz,Baier:2006pt,Renk:2006sx}.}   \cite{Sadofyev:2021ohn,Barata:2022krd,Andres:2022ndd,Barata:2023qds,Kuzmin:2023hko,Kuzmin:2024smy}. However, many of the transport-based approaches used in jet quenching phenomenology, see e.g.~\cite{Kurkela:2021ctp,Boguslavski:2023fdm,Boguslavski:2023waw,Boguslavski:2024kbd,Kurkela:2018wud,Schlichting:2020lef,Mehtar-Tani:2022zwf,Du:2020zqg,Du:2020dvp,Du:2023izb,BarreraCabodevila:2022jhi,Cabodevila:2023htm,Zhou:2024ysb}, are theoretically based on the
kinetic theory of QCD~\cite{Arnold:2002zm,Arnold:2000dr,Caron-Huot:2010qjx}, which accounts only for local interactions and thus misses effects beyond Eq.~\eqref{eq:diff_simple}. On the other hand, while simulation frameworks such as JEWEL \cite{Zapp:2008gi,Zapp:2012ak}, LBT \cite{He:2015pra,Cao:2016gvr}, or LIDO \cite{Ke:2018tsh} rely on field-theory-based descriptions of high-$p_t$ probes, which have been recently extended to account for the medium structure and evolution \cite{Sadofyev:2021ohn,Barata:2022krd,Andres:2022ndd,Barata:2023qds,Kuzmin:2023hko,Kuzmin:2024smy,Hauksson:2021okc,Hauksson:2023tze,Barata:2024bqp}, such developments have not been fully integrated into any simulations yet.\footnote{See~\cite{He:2020iow,Antiporda:2021hpk,Barata:2023zqg} for some recent developments in this direction.} 
In turn, similar effects due to the hydrodynamic gradients arise in holographic considerations, see e.g. the discussion in \cite{Lekaveckas:2013lha,Rajagopal:2015roa,Sadofyev:2015hxa,Reiten:2019fta,Arefeva:2020jvo}, and should be integrated into the Hybrid model \cite{Casalderrey-Solana:2014bpa,Casalderrey-Solana:2015vaa,Casalderrey-Solana:2016jvj}. Thus, when non-trivial background profiles are considered, even the functional structure of the evolution equations can be modified, and solely incorporating spacetime-dependent transport coefficients into an AMY-based picture only partially accounts for the effects induced by the medium anisotropy. We note that most of these limitations remain in place during the earlier stages of jet evolution, where a non-QGP, anisotropic, out-of-equilibrium background is present, see e.g.~\cite{Ipp:2020mjc,Ipp:2020nfu,Carrington:2021dvw,Avramescu:2023qvv,Boguslavski:2023alu,Boguslavski:2023waw,Backfried:2024rub,Du:2023izb,Barata:2024xwy,Pandey:2023dzz,Avramescu:2024poa,Avramescu:2024xts} for recent discussions on hard probe transport in these initial stages.

Recently, some of us showed that within a hydrodynamic gradient expansion of the medium density, the standard diffusion equation in Eq.~\eqref{eq:diff_simple} receives non-trivial corrections starting at second order in the gradient expansion~\cite{Barata:2022utc}. Thus, while at first order in gradients the above prescription using AMY/Boltzmann transport with a spacetime-dependent $\hat q$ is valid \cite{Fu:2022idl}, beyond this order, the master equations themselves are modified \cite{Barata:2022utc}. In the present case, as illustrated below, the novel collisional kernel results from non-local interactions between the parton and the medium constituents. Its explicit form was derived from QFT considerations, extending beyond the AMY theory description of the single parton evolution. It is worth noting that similar non-local transport equations may also arise in the context of spin-hydrodynamics and spin-kinetic theories~\cite{Weickgenannt:2020aaf,Sheng:2021kfc,Weickgenannt:2022qvh,Weickgenannt:2021cuo,Sheng:2020oqs,Weickgenannt:2024ibf}, where they are again derived from a QFT perspective, which allows going beyond classical transport.

Despite progress in deriving a new master equation for the evolution of a single parton in the medium, its form was too complex to fully determine the phenomenological importance of matter gradient corrections. The main goal of this paper is thus to complete this step and provide a thorough numerical study of the new transport equation. To this end, we construct a simplified model for the spatial 
profile of the medium density and solve for the evolution of the spatial and momentum distributions characterizing the hard parton. We compare the solutions of the full evolution equation provided in~\cite{Barata:2022utc} with simpler forms of transport to assess the relevance of gradient corrections. As we show below, we find that the azimuthal structure of the distribution relative to the propagation direction of the hard parton is significantly modified by the medium's structure. Thus, to use jets as a tomographic tool for the QGP, it is essential to account for modifications to the functional form of the transport master equations.

The paper is organized as follows: in section~\ref{sec:theory}, we review the main results from~\cite{Barata:2022utc} and detail how they can be properly conditioned for numerical implementation. In section~\ref{sec:numerics}, we present a detailed study of the master evolution equation in different scenarios, including the cases of a static inhomogeneous medium and a flowing medium governed by a hydrodynamic evolution. Finally, in section~\ref{sec:conclusion}, we discuss the main results of our study. Additional results and details are provided in the Appendix.

\section{Theoretical setup}\label{sec:theory}

We start by recalling the main result from~\cite{Barata:2022utc}: the evolution equation for the Wigner function, $W(\Y,\pv)$, associated with a single hard parton propagating in the presence of a QCD background with a non-uniform spatial density, $\rho(\Y)$. At second order in the spatial gradient expansion of $\rho(\Y)$, it can be compactly written as
\begin{align}\label{eq:main_result}
&\left(\pa_L+\frac{\pv\cdot\tvec{\na}_\Y}{E}-\frac{\hat{q}(\tvec{Y})}{4}\pa_\p^2\right) W(\Y,\pv)\notag\\
&\hspace{3cm}=\na_i\na_j\rho(\Y)\,\int_{\q}\left[\kappa \frac{\pa^2}{\pa p_i\pa p_j} \d^{(2)}(\qv)-V_{ij}(\q)\right]W(\Y,\pv-\qv)\, ,
\end{align}
where 
$\kappa    =  \frac{\pi^2}{2} C_F\int_{\q} (v(\q^2))^2 $ with $C_F$ being the quadratic Casimir of the fundamental representation, and we adopt the shorthand notations $\int_\q = \int 
\frac{d^2\q}{(2\pi )^2}   $ and $\int_\Y = \int d^2\Y   $. Note that for the evolution of a high-$p_t$ parton, all the non-trivial dynamics occur in the transverse plane relative to the propagation, so only transverse gradients of $\rho(\Y)$ emerge. Thus, we use bold symbols to denote the transverse momentum of the parton $\pv$ and its spatial location in the medium $\Y$, while $E$ represents its total energy. The potentials $v(\q^2)$ describe the  
interactions between the hard probe and the medium; here, we assume they follow the Gyulassy-Wang (GW) model~\cite{Gyulassy:1993hr}:
\begin{align}
   v(\q^2)=\frac{-g^2}{\q^2+m_D^2} \, , 
\end{align}
where $m_D$ is the Debye screening mass.
For this model, it follows that $\kappa    =  g^4\pi C_F/8m_D^2$.
The momentum distribution and the diffusive evolution introduced in Eq.~\eqref{eq:diff_simple} can be directly recovered by integrating Eq.~\eqref{eq:main_result} over space, i.e. $\cP(\pv)=\int_\Y W(\Y,\pv)$, in the limit of a homogeneous medium. 

The right hand side of Eq.~\eqref{eq:main_result} corresponds to a non-local collisional kernel that modifies the form of the diffusion equation. More explicitly, the interactions are governed by
\begin{align}
&V_{ij}(\qv)=\frac{C}{2}\Bigg(\Big\{2q_i q_j \left[vv^{\prime\prime}-v^\prime v^\prime\right]+vv^\prime\delta_{ij}\Big\}\notag\\
&\hspace{4cm}-(2\pi)^2\d^{(2)}(\qv)\int_\lv\Big\{2l_i l_j \left[v v^{\prime\prime}-v^\prime v^\prime\right]+vv^\prime\delta_{ij}\Big\}\Bigg) \, ,
\end{align}
where %we used 
$h'(\q^2)\equiv \partial_{\q^2} h(\q^2)$ and $C=\frac{C_F}{2N_c}$, assuming the sources and projectiles are in the fundamental representation. Furthermore, one should understand the space-dependent jet quenching parameter as being expanded in gradients, i.e.  $\hat q(\Y)= \hat q + \Y \cdot \nabla \hat q + \frac{1}{2} Y_i Y_j\nabla_i \nabla_j \hat q$. Since we only include gradients of the density but not of the Debye mass, it follows that $\rho \tvec{\na}\hat{q} = \hat{q} \tvec{\na} \rho$, and the homogeneous $\hat q$ is defined through the homogeneous $\rho$. The relation between these homogeneous terms can be made explicit in the GW model. For that we introduce the so-called dipole potential
\begin{align}
\mathcal V(\tvec{q})\equiv -C\,\r\left(\left|v(\q^2)\right|^2-(2\pi)^2\delta^{(2)}(\tvec{q})\int_{\tvec{l}}\,\left|v(\l^2)\right|^2\right)\,,    
\end{align}
which, in the diffusion picture assumed above, takes the form
\begin{align}\label{eq:V_harm}
\mathcal{V}(\y)\equiv \int_\q e^{i\q \cdot \y} \,  \mathcal{V}(\q)\approx \frac{\hat q}{4}    \y^2\, ,
\end{align}
where $\hat{q}\equiv 4\pi\,C_F\, \alpha_s^2  \rho \log \frac{Q^2}{m_D^2}  $, with $Q$ being a free large momentum scale ubiquitous to the harmonic (diffusive) approximation being employed, and $\alpha_s=g^2/(4\pi)$. For what follows, this scale will play no role and can be thought of as being absorbed into the definition of the homogeneous coefficients.

The non-local form of the right hand side of Eq.~\eqref{eq:main_result} makes it challenging for numerical evaluation. However, in the diffusion approximation, one can further simplify the collisional kernel by performing a power expansion in the exchanged momentum $\q$, assuming the total momentum $\p$ to be much larger. Notice that to obtain the diffusion equation
on the left hand side, a similar exercise has already been performed, as exemplified in Eq.~\eqref{eq:V_harm}. After expanding to the lowest non-trivial order, we obtain the localized evolution equation
\begin{align}
&\left(\pa_L+\frac{\pv\cdot\tvec{\na}_\Y}{E}-\frac{\hat{q} + \Y \cdot  \tvec{\na} \hat q+ \frac{1}{2} Y_i Y_j  \na_i \na_j \hat q }{4}\pa_\p^2\right) W(\Y,\pv)\notag\\
&\hspace{1cm}=\na_i\na_j\rho\,\left[\frac{\kappa}{(2\pi)^2} \frac{\pa^2}{\pa p_i\pa p_j}-\int_{\q}V_{ij}(\q)\q\cdot\frac{\pa}{\pa \p}-\frac{1}{2}\int_{\q}V_{ij}(\q)q_aq_b\frac{\pa^2}{\pa p_a\pa p_b}\right]W(\Y,\pv)\, ,
\end{align}
which can be further simplified for the specific potential, satisfying
\begin{align}
&\int_{\q}V_{ij}(\q) =\int_{\q}V_{ij}(\q) q_a = 0 \, ,\notag\\
&\int_{\q}V_{ij}(\q) q_a q_b = \frac{C_Fg^4}{8}(\d_{ij}\d_{ab}+\d_{ia}\d_{jb}+\d_{ib}\d_{ja})\int_{\q}q^4\left(vv^{\prime\prime}-v^\prime v^\prime\right)+\frac{C_Fg^4}{4}\d_{ij}\d_{ab}\int_{\q}q^2vv^\prime\notag\\
&\hspace{2.3cm}=\frac{C_Fg^4}{96\pi m_D^2}(\d_{ia}\d_{jb}+\d_{ib}\d_{ja}-2\d_{ij}\d_{ab}) \, .
\end{align}
Combining all these elements, we find that in the small-angle scattering limit, Eq.~\eqref{eq:main_result} takes the simple form
\begin{align}\label{eq:final_evol_eq}
&\left(\pa_L+\frac{\pv\cdot\tvec{\na}_\Y}{E}-\frac{\hat{q} + \Y \cdot  \tvec{\na} \hat q+ \frac{1}{2} Y_i Y_j  \na_i \na_j \hat q }{4}\pa_\p^2\right) W(\Y,\pv)\notag\\
&\hspace{5cm}=\frac{ \alpha_s^2 \pi  C_F}{6 m_D^2} \na_i\na_j\rho\,\left[2\d_{ia}\d_{jb}+\d_{ij}\d_{ab}\right]\frac{\pa^2}{\pa p_a\pa p_b}W(\Y,\pv)\, .
\end{align}
The expansion term $\pv\cdot\tvec{\na}_\Y/E$ leads to a negligible change in the evolution of the distribution due to the small velocities $p_x, p_y \ll E$. This equation can be numerically evaluated in an efficient manner by discretizing over the momentum and position space lattices. Details of the implementation are provided in the next section, with additional information given in Appendix~\ref{app:lattice}.

\section{Numerical results}\label{sec:numerics}
In this section, we examine the solutions to the master equation Eq.~\eqref{eq:final_evol_eq} in two idealized scenarios. In the first case, the medium is assumed to be static, with an isotropic but inhomogeneous density profile. We then study the evolution of a single parton initially produced at different points in the medium. An equivalent example for an anisotropic and inhomogeneous system is provided in Appendix~\ref{app:extra_static}. Note that, since we initiate the parton at different spatial locations, the matter appears locally anisotropic to the probe (except at the origin). Thus, there is no qualitative difference between these two cases, as demonstrated numerically. In the second part of this section, we study the case where the probe propagates through matter that 
flows transversely to its propagation direction, which leads to a generalization of 
Eq.~\eqref{eq:final_evol_eq}.

\subsection{Static medium}

We first consider a simple setup, where the matter is static and has a non-trivial geometry:
\begin{align}\label{eq:qhat_model1}
    \hat{q}(\Y)=\hat{q}_0(\Y)\ln\bigg(\frac{Q^2(\Y)}{m_D^2(\Y)}\bigg),~~~
    \hat{q}_0(\Y)=4\pi \alpha_s^2 C_F\rho(\Y
    ),~~~
    \rho(\Y)=\rho_0\exp(-A_{ij}Y_iY_j) \, ,
\end{align}
where we set 
$Q^2(\Y)=E\, m_D(\Y)$, $m_D(\Y)$ is the position-dependent Debye mass, $\rho(\Y)$ is assumed to be exponentially decaying, and $\alpha_s$ is taken to be fixed.

After inserting Eq.~\eqref{eq:qhat_model1} into Eq.~\eqref{eq:final_evol_eq}, the master evolution equation for the Wigner function can be written as 
\begin{align}
\label{eq:numerics_dif_eq}
\left(\pa_L+\frac{\pv\cdot\tvec{\na}_\Y}{E}-D_{ab}\frac{\pa^2}{\pa p_a\pa p_b}\right)W(\Y,\p)=0\,,
\end{align}
where the diffusion matrix $D_{ab}$ for this medium profile $\rho(\Y)$ is explicitly given by 
\begin{align}
\label{eq:diff_coeff}
D_{ab}(\Y)&=\frac{\hat{q}(\Y)}{4}\d_{ab}+\frac{\alpha_s^2 \pi C_F}{3 m_D^2(\Y)}\rho(\Y)\left[4A_{ak}A_{bl}Y_kY_l-2A_{ab}+2\d_{ab}A_{ik}A_{il}Y_k Y_l-\d_{ab}A_{ii}\right]\,.
\end{align}
This diffusion coefficient consists of two components: one represented as $\hat{q}(\Y)\delta_{ab}/4$, and the remaining terms, which cannot be absorbed into the quenching parameter $\hat{q}(\Y)$ and will be referred to as the gradient correction terms. For a medium profile without off-diagonal terms, where $A_{ab}=\delta_{ab}\mu_{a}^2$, one has
\begin{align}
D_{xx}(\Y)&= \frac{\hat q(\Y) }{4}+ \frac{\alpha_s^2 \pi C_F }{3m_D^2(\Y)} \rho(\Y)\left(4\mu_x^4 Y_x^2+2(\mu_x^4Y^2_x+\mu_y^4Y_y^2)-3\mu_x^2-\mu_y^2\right) \, ,\nn
D_{xy}(\Y)&=\frac{4 \pi \alpha_s^2 C_F}{3 m_D^2(\Y)}\,\rho(\Y)\,\mu_x^2\mu_y^2 Y_x Y_y \, .
\end{align}
In a locally thermal QCD plasma with a temperature $T(\Y)$ profile, the Debye mass is position-dependent and can be related to the temperature by~\cite{Kapusta:2006pm}
\begin{align}
    m_D^2(\Y)
    =g^2\bigg(1+\frac{N_f}{6}\bigg)T^2(\Y)= \frac{3}{2}g^2T^2(\Y) \, ,
\end{align}
with $N_f=3$ flavors.\footnote{Note that in this work, we do not include gradients of $\tvec{\na} m_D$. As a result, there are missing contributions to the master equation, since both the Debye mass and the density are only function of $T$.} The position-dependent temperature and energy density are evaluated from the number density via Landau matching 
\begin{align}
\label{eq:landau_matching}
    \nonumber
    \rho(\Y)&=\bigg(\nu_g\frac{\zeta(3)}{\pi^2}+\nu_q N_f\frac{3\zeta(3)}{2\pi^2}\bigg)T^3(\Y)
    \simeq\bigg(1.9487+1.0961 N_f\bigg)T^3(\Y)\, ,\\
    e(\Y)&=\bigg(\nu_g\frac{\pi^2}{30}+\nu_q N_f\frac{7\pi^2}{120}\bigg)T^4(\Y)
    \simeq\bigg(5.2638+3.4544 N_f\bigg)T^4(\Y)\, .
\end{align}
where $\nu_g=16$ and $\nu_q=6$ are degeneracy factors for gluons and quarks respectively, and $\zeta(3)\simeq 1.2$ is the Riemann zeta function.
Notice that for Eq.~\eqref{eq:numerics_dif_eq} to be numerically stable--that is, to avoid exponentially diverging solutions--one has to require that $D_{xx}>0$ and $D_{yy}>0$. For the initial condition of the hard parton, we assume a Gaussian form 
\begin{align}\label{eq:W_0}
W(\Y,\p,L=0)=f(\Y)e^{-p^2/b^2} \, .  
\end{align}
where $b$ is a parameter that determines the dispersion in momentum space. The functional form of the positional distribution is not critical, as it factors out of the results, assuming that $W(\Y,\p,0)\equiv f(\Y) g(\p)$, and at high energy, where sub-eikonal terms can be neglected in Eq.~\eqref{eq:numerics_dif_eq}.

Using the freedom to define energy units, we choose the typical saturation scale $Q_s$ as the reference and express all quantities in terms of $Q_s$. For example, $E=\varepsilon $ means $ E=\varepsilon Q_s$, regardless of the value of $Q_s$ (typically 1\,GeV). In numerical calculations, we do not associate the setup with any specific collisional system. Instead, we assume reasonable parameter values within a realistic range for heavy-ion collisions.
We choose a medium profile with $\rho_0=5$, $\mu_x=\mu_y=1$, and coupling $g=1$ as the default. In addition to this isotropic case, we also consider an anisotropic medium %profile 
with $\mu_x=1$, $\mu_y=2$, see Fig.~\ref{fig:medium_pic} in Appendix~\ref{app:extra_static} for illustrations. The hard parton profile is assumed to have a total energy of $E=1000$, and we take $b=1$ in Eq.~\eqref{eq:W_0}. To justify the small-gradient expansion of the medium profile and hydrodynamic description, we require that the parametric relation $\nabla T/T^2\sim \mu/T\lesssim 1$ holds within the considered region in $\Y$.~\footnote{According to Eq.~(\ref{eq:landau_matching}), we have $T\sim\frac{\rho_0}{5.237}\exp(-\frac{\mu^2Y^2}{3})\sim \exp(-\frac{\mu^2Y^2}{3})$ and $|\nabla T/T^2|\sim|\frac{2\mu^2Y}{3}|\exp(\frac{\mu^2Y^2}{3})$. In fact, for illustrative purposes, the simulations (in the static medium case) explore a region where the gradients are larger than this bound.}
The evolution time of the hard parton is characterized by the distance $L$ traveled in the longitudinal direction, which is perpendicular to the transverse plane where the gradient corrections contribute.

\begin{figure}[h!]
    \centering
    \includegraphics[width=0.49\textwidth]{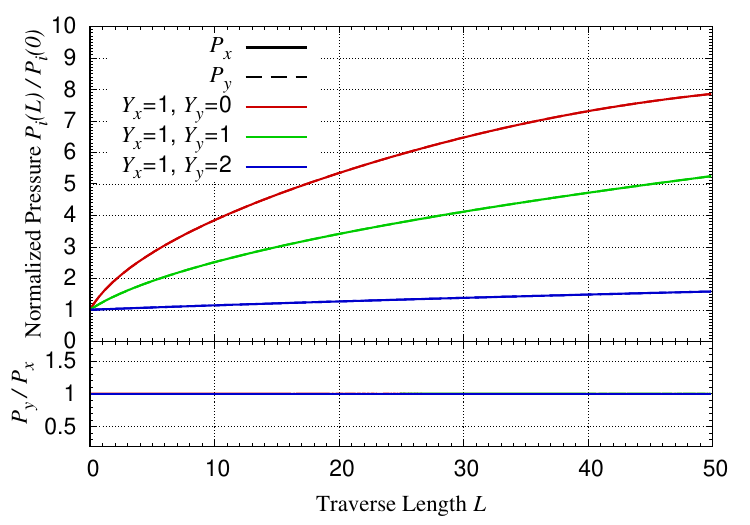}
    \includegraphics[width=0.49\textwidth]{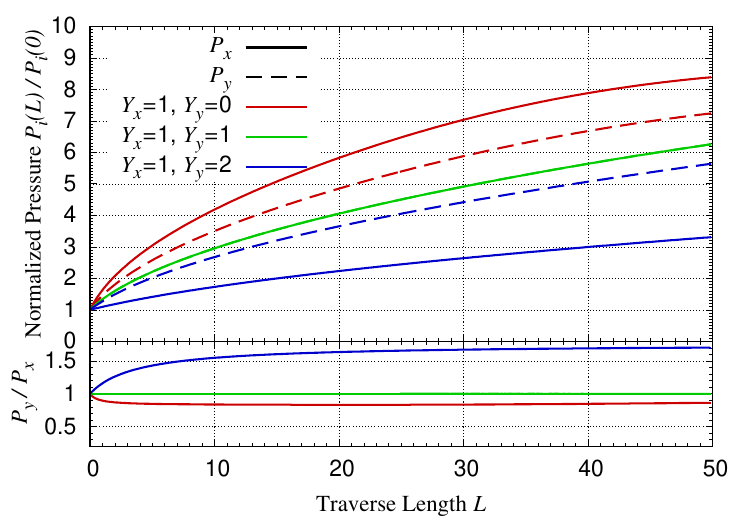}
    \caption{Transverse pressures $P_x$, $P_y$ associated with the evolution of a single hard parton (quark) in the medium. Solid (dashed) lines show the pressure along the $x$ ($y$) direction. The different colors indicate the spatial point where pressure is measured along the line $Y_x=1$ from the center $Y_y=0$ to the edge $Y_y=2$ of medium (see medium profile in Fig.~\ref{fig:medium_pic}).
    \textbf{Left:} Results obtained for the case where all gradient corrections that can not be absorbed into $\hat q$ are neglected. \textbf{Right:} Results including all the second-order gradient corrections to the evolution.}
    \label{fig:pressure_Y=X}
\end{figure}

We first extract the time evolution of the transverse pressure along the $x$ and $y$ directions associated with the hard parton, defined as
\begin{align}
    P_i(\Y,t)=\int_{\p} \, \frac{p_i^2}{\sqrt{\p^2}}W(\Y,\p,t) \,.
\end{align}
where $i=x,y$. In Fig.~\ref{fig:pressure_Y=X}, we plot the pressure normalized to its initial time value, $P_{x}(\Y,L)/P_{x}(\Y,0)$ and $P_{y}(\Y,L)/P_{y}(\Y,0)$, as well as the pressure ratio $P_{y}(\Y,L)/P_{x}(\Y,L)$, both with and without gradient corrections. The left panel in Fig.~\ref{fig:pressure_Y=X} shows the results for the case where gradient corrections that cannot be absorbed in $\hat q(\Y)$ are neglected, reducing Eq.~\eqref{eq:final_evol_eq} to the diffusive case in Eq.~\eqref{eq:diff_simple} with a spatial dependent $\hat q(\Y)$. The right panel displays the solution obtained by solving Eq.~\eqref{eq:final_evol_eq} in full. We note that the parton energy, $E=1000$, is large, ensuring that the transverse velocity $\v\ll 1$, allowing us to neglect the term involving spatial derivatives.

\begin{figure}[h!]
    \centering
    \includegraphics[width=0.90\textwidth]{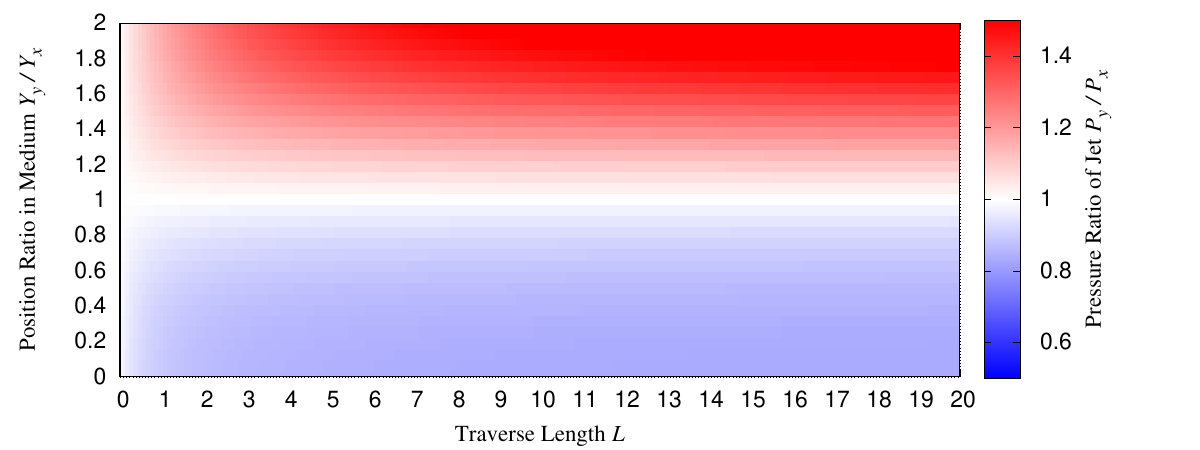}
    \caption{Pressure ratio $P_y/P_x$ evolution over the parton path length as a function of parton distance ratio $Y_y/Y_x$ (with $Y_x=1$) to the center of the medium}
    \label{fig:pressure_Y=X_v2}
\end{figure}

Focusing first on the left panel of Fig~\ref{fig:pressure_Y=X}, one finds the expected behavior: at any spatial location, the pressures in the $x$ and $y$ directions are identical, meaning the dashed and solid curves overlap for each color. Furthermore, at the spatial location with the highest medium density (red curve), one observes the fastest pressure growth over time. Conversely, in the opposite case (blue curve), the time evolution is slow. When the full second-order gradient corrections are included (right panel of Fig.~\ref{fig:pressure_Y=X}), the pattern is qualitatively different. Firstly, when $x\neq y$, the dashed and solid lines no longer overlap. 
This indicates that even in a medium with a completely isotropic profile, gradient effects can deform the particle distribution of the hard parton, pushing it more in the direction of the largest gradient. Notice that such a spatially dependent effect cannot be captured solely by making the transport coefficient space-dependent. The separation of the curves is most pronounced for the largest separation in $\Y$, while for $Y_x=Y_y$ (green curve) they overlap exactly. 
For both plots, we show $P_y/P_x$ below, with these ratios tending to approach a constant upper or lower limit at late times. Importantly, in the right panel of Fig~\ref{fig:pressure_Y=X}, the modifications induced by the new collisional kernel can be substantial, leading to an $\mathcal{O}(20\%-60\%)$ deviation.

To complement Fig.~\ref{fig:pressure_Y=X}, we present the pressure ratio as a function of $L$ in Fig.~\ref{fig:pressure_Y=X_v2} for various initial position, expressed in terms of $Y_y/Y_x$ with $Y_x=1$. Again, starting from an isotropic configuration, we see that the new scattering kernel introduces an asymmetry between the $x$ and $y$ directions, due to the presence of density gradients.

\begin{figure}[!h]
    \centering
    \includegraphics[width=0.86\textwidth]{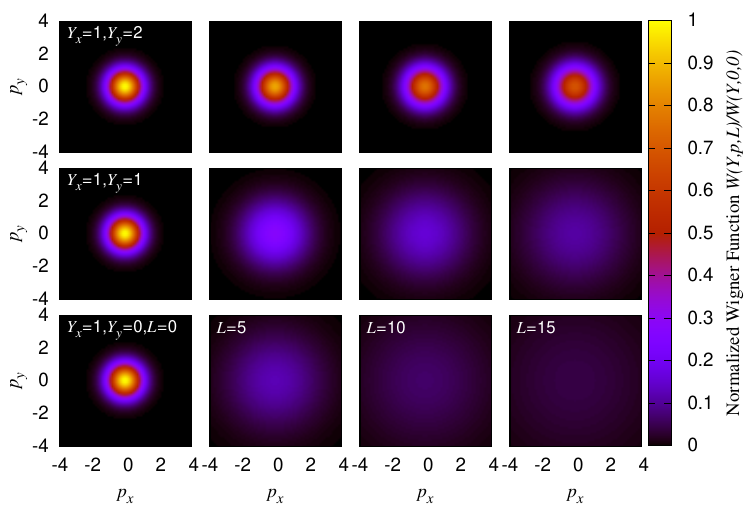}
    \includegraphics[width=0.86\textwidth]{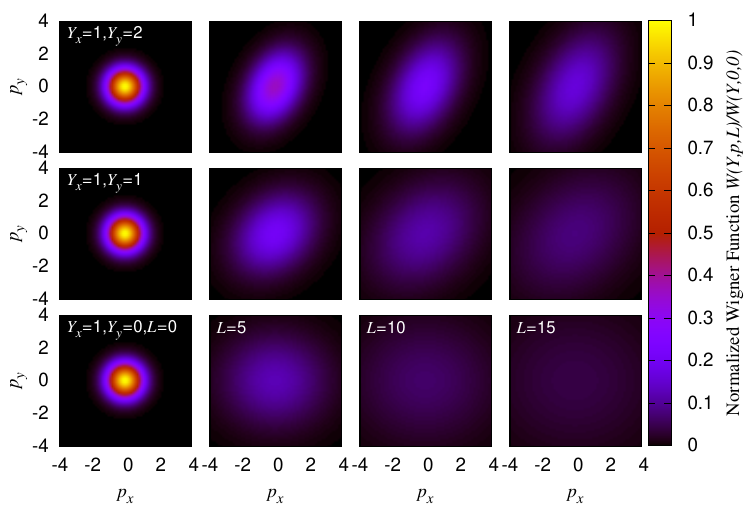}
    \caption{Heat maps of the hard parton (quark) momentum distribution along the line $Y_x=1$ at points $Y_y=0,1,2$, and at different lengths $L$=0, 5, 10, 15, without (upper panel) and with (lower panel) the gradient correction to the diffusion. The 2D distributions have been normalized to 1 at the peak $p_x$=0, $p_y$=0 at $L$=0 for each position $\Y$.}
    \label{fig:strobo_RHS_1}
\end{figure}

\begin{figure}[!h]
    \centering
    \includegraphics[width=0.86\textwidth]{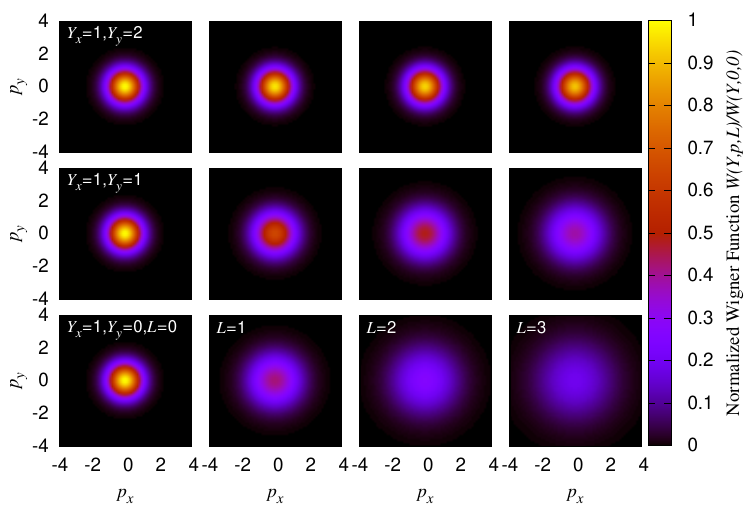}
    \includegraphics[width=0.86\textwidth]{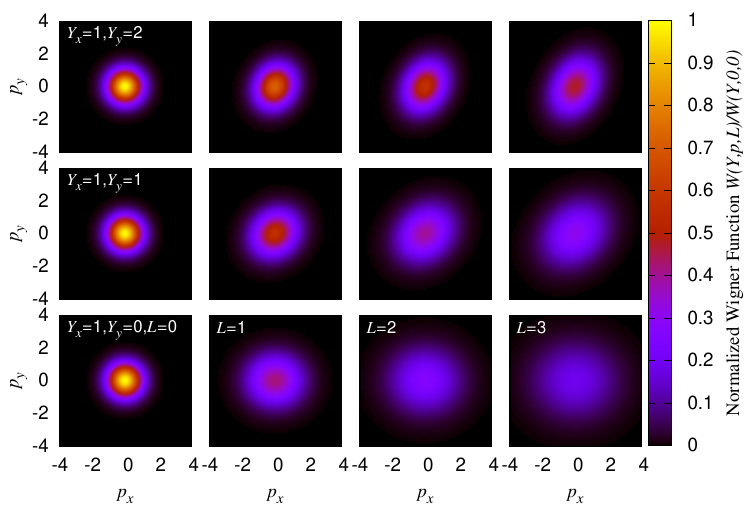}
    \caption{Same results as in Fig.~\ref{fig:strobo_RHS_1}, but now for $L=0,1,2,3$. The evolution is without (upper panel) and with (lower panel) the gradient correction to the diffusion.}
    \label{fig:strobo_RHS_2}
\end{figure}

To better illustrate the modifications induced by the gradients, we present the stroboscopic evolution of $W(\Y,\p,t)$, measured at the spatial point $Y_x=1$ and $Y_y=0,1,2$ in Fig.~\ref{fig:strobo_RHS_1}. The upper (lower) panel in Fig.~\ref{fig:strobo_RHS_1} displays the result without (with) gradient corrections, corresponding to the colored curves in the left (right) panel of Fig.~\ref{fig:pressure_Y=X}. 
As evident from direct inspection, the evolution considering only the corrections in $\hat q(\Y)$ (upper in Fig.~\ref{fig:strobo_RHS_1}) results in a perfectly isotropic distribution. 
This exercise, considering the evolution governed by Eq.~\eqref{eq:final_evol_eq} (lower in Fig.~\ref{fig:strobo_RHS_1}), reveals a noticeable effect: the momentum distribution develops a non-trivial azimuthal shape driven by the matter gradients.
Thus, by comparing these results, it is reasonable to conclude that properly describing the azimuthal structure of jets evolving in the QGP requires going beyond standard Boltzmann transport. 
In Fig.~\ref{fig:strobo_RHS_2}, we also present the detailed early-time evolution of the Wigner distribution within the same setup, to highlight the emergence of the anisotropic structure.

\begin{figure}[h!]
    \centering
    \includegraphics[width=0.7\textwidth]{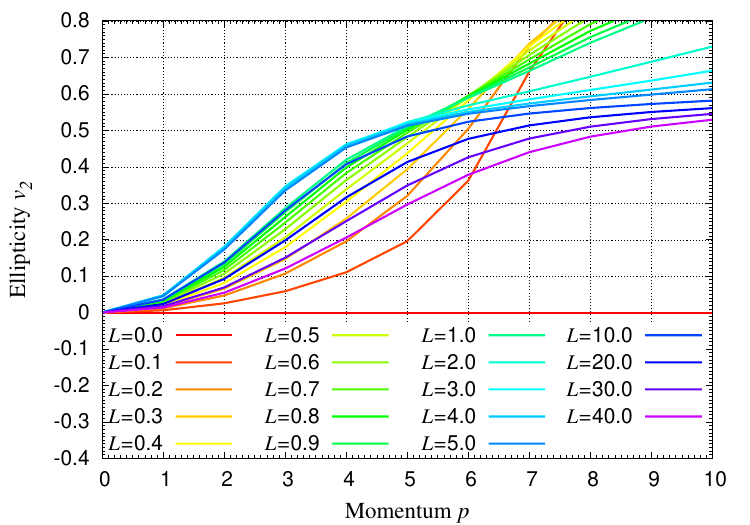}
    \caption{Ellipticity for a jet located at $Y_x=1$, $Y_y=2$, using the full evolution equation. The equivalent result with gradient corrections only coming through $\hat q$ vanishes for all $L$.}
    \label{fig:v_2}
\end{figure}

Finally, to better quantify the degree of anisotropy in Fig.~\ref{fig:strobo_RHS_1} and Fig.~\ref{fig:strobo_RHS_2}, we compute the leading non-vanishing Fourier harmonic in Fig.~\ref{fig:v_2}, where we take the $y$-axis as the reference direction.
We present explicit results only for the solutions obtained using the full master equation, while ellipticity vanishes exactly when considering only gradient effects.
The same holds for all odd harmonics, as expected. As can be seen, the presence of the gradients leads to a substantial elliptical modulation of the distribution, which approaches a nearly constant behavior at late times.

\subsection{Flowing medium}
We now consider the case where the underlying medium flows transversely to the jet, i.e. the velocity field of the medium has the form $\vec{v}=(\textbf{v},0)=(v_x,v_y,0)$. The rationale for choosing this setup is that with a purely transverse flow, one only needs to solve $2+1$-d hydrodynamics, which is significantly less complex than the $3+1$-d case. Nonetheless, this setup still allows for a qualitative understanding of the effects of flow on jet transport, which is the aim of this work.

In the following simulation, we consider two cases: one with a constant flow as a toy model and the other with a hydrodynamic flow solved from hydrodynamic equations in Minkowski space
\begin{align}
    D_{\mu}T^{\mu\nu}=0\,.
\end{align}
The energy-momentum tensor $T^{\mu\nu}(\Y)$ in hydrodynamic theory can be initialized by converting the number density $\rho(\Y)$ 
into the energy density and pressure. This is done using the conformal equation of state (EoS), $e=3p$, and applying Landau matching at each position $\Y$ in Eq.~(\ref{eq:landau_matching}).
Details of the numerical strategy used to obtain the hydrodynamic simulation can be found in the Appendix~\ref{app:hydro}.
In summary, we implemented the Kurganov-Tadmor scheme~\cite{Kurganov:1999,Bazow:2016yra,Du:2019obx} as a finite difference method in the hydrodynamic code. We further validate the code by comparing its numerical results with the analytical solution for Gubser flow~\cite{Gubser:2010ze,Gubser:2010ui,Nopoush:2014qba,Pang:2014ipa} in Milne/de Sitter space.

In addition to the more complex background, the master transport equation must also be updated to account for the medium's flow. Building on the considerations in~\cite{Barata:2022utc} and incorporating new theoretical elements describing jet propagation in flowing and structured matter~\cite{Kuzmin:2023hko}, one can show that for a two-dimensional transverse flow field $\tvec{v}$, the Wigner function satisfies:
\begin{align}
\label{eq:equation_with_flow}
&\left(\pa_L-\tvec{v}\cdot\tvec{\na}_\Y+\frac{\pv\cdot\tvec{\na}_\Y}{E}-\frac{\hat{q} \left(1-L\tvec{\na}\cdot\tvec{v}\right) + \Y \cdot  \tvec{\na} \hat q+ \frac{1}{2} Y_i Y_j  \na_i \na_j \hat q }{4}\pa_\p^2\right) W(\Y,\pv)\notag\\
&\hspace{1cm}=\na_i\na_j\rho\,\left[\frac{\kappa}{(2\pi)^2} \frac{\pa^2}{\pa p_i\pa p_j}-\int_{\q}V_{ij}(\q)\q\cdot\frac{\pa}{\pa \p}-\frac{1}{2}\int_{\q}V_{ij}(\q)q_aq_b\frac{\pa^2}{\pa p_a\pa p_b}\right]W(\Y,\pv)\, ,
\end{align}
which can be shown to preserve the normalization of the Wigner function up to the given order in the gradient expansion. Thus, the flow enters in two ways: as a new convection term—similar to the sub-eikonal factor already present at eikonal order—and as a multiplicative \textit{shift} to the bare $\hat q$. The latter arises from the fact that while the probe propagates through the medium, the medium itself is also moving \cite{Kuzmin:2023hko}. 

In what follows, we first explore the case where $\tvec{v}$ is a constant vector, turning on and off the right hand side term of the evolution equation for the jet. This simple background setup allows us to gauge the effects of the novel terms in transport with flow. Secondly, we determine $\tvec{v}$ by solving the hydrodynamic equations for a given initial energy profile. The initial profile is chosen to fit within the simulation lattice while ensuring that pressure gradients remain small enough for Eq.~\eqref{eq:equation_with_flow} to hold.

\subsubsection{Constant flow}
\begin{figure}[]
    \centering
    \includegraphics[width=0.86\textwidth]{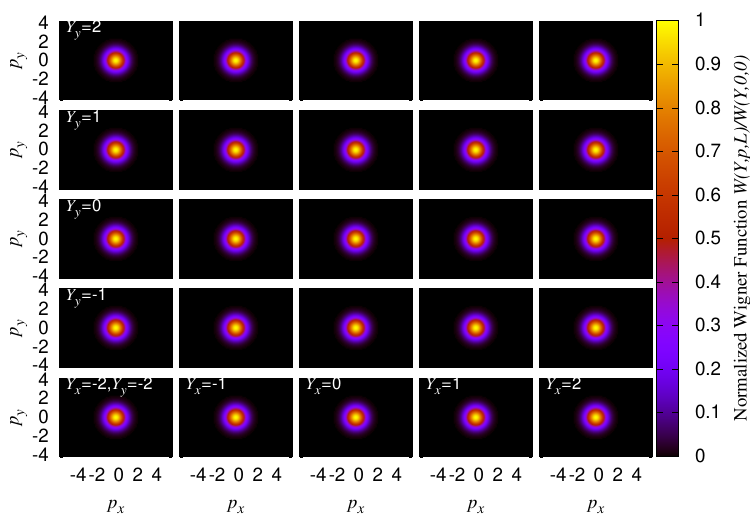}
    \includegraphics[width=0.86\textwidth]{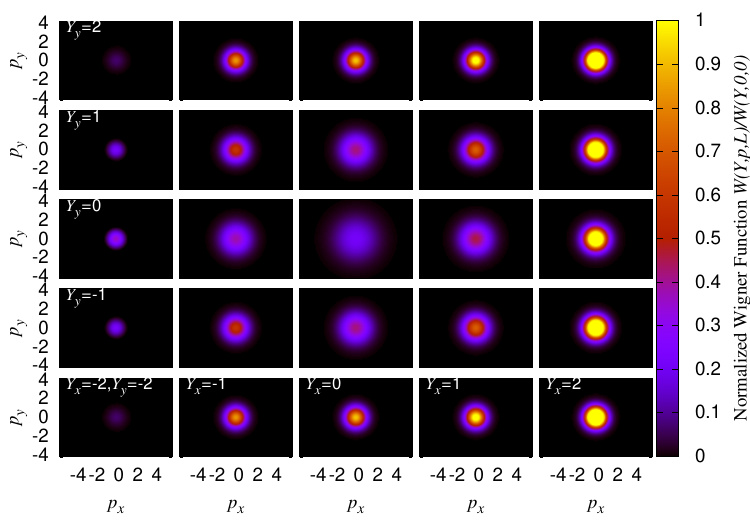}
    \caption{Heatmap for the  Wigner function at various positions. Both expansion and flow effects are included, neglecting the gradient correction term, i.e. the evolution follows the left hand side of Eq.~(\ref{eq:equation_with_flow}). 
    The flow is assumed to be constant towards the $x$ direction: $v_x=0.1$. The upper and lowers plots correspond to $L=0$ and $L=1$, respectively.}
    \label{fig:strobo_Flow_constant1}
\end{figure}

\begin{figure}[]
    \centering
    \includegraphics[width=0.86\textwidth]{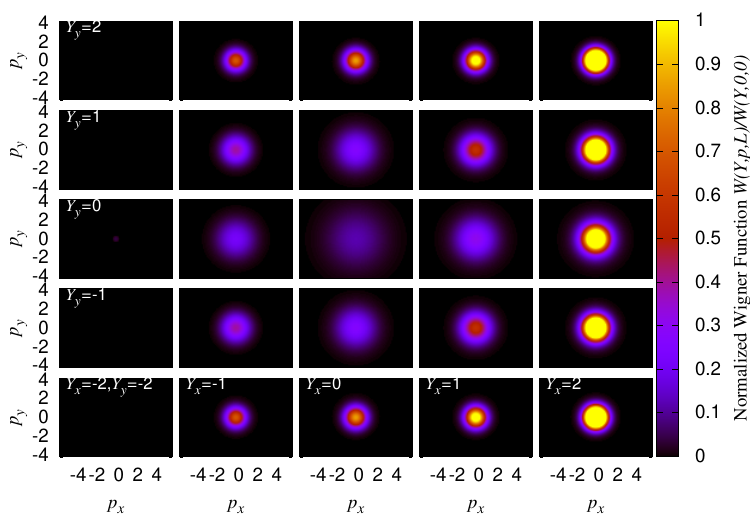}
    \includegraphics[width=0.86\textwidth]{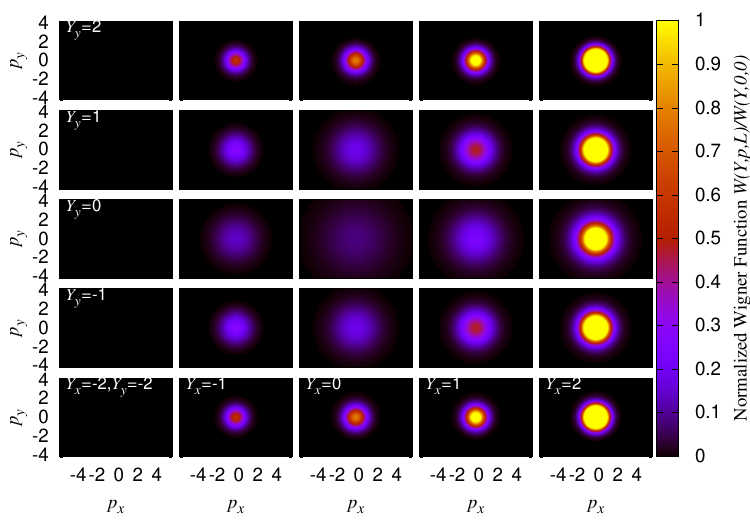}
    \caption{Same results as in Fig.~\ref{fig:strobo_RHS_Flow_constant1}. The upper and lowers plots correspond to $L=2$ and $L=3$, respectively.}
    \label{fig:strobo_Flow_constant2}
\end{figure}

In Figs.~\ref{fig:strobo_Flow_constant1} and \ref{fig:strobo_Flow_constant2}, we show the evolution of the Wigner distribution at a series of positions as a function of momentum for $L=0,1,2,3$, excluding the gradients terms on the right hand side of Eq.~\eqref{eq:equation_with_flow}. The flow vector has the form $\tvec{v}=(v_x=0.1,v_y=0)$, pointing from left to right in the figures. The initial condition for the jet remains the same as in the previous simulations.

A direct analysis of the results shows that as time progresses, the initially isotropic and homogeneous solution at $L=0$ evolves into one where more modes are excited along the flow direction. Physically, this result is reasonable, as one expects the Wigner distribution density to increase along the underlying flow field. 
More importantly, while the Wigner distribution becomes anisotropic under evolution with gradients and flow, for a fixed impact parameter (at non-central positions), the momentum-space distribution remains fairly isotropic.
This is analogous to the behavior observed in the previous section for non-flowing matter.

\begin{figure}[]
    \centering
    \includegraphics[width=0.86\textwidth]{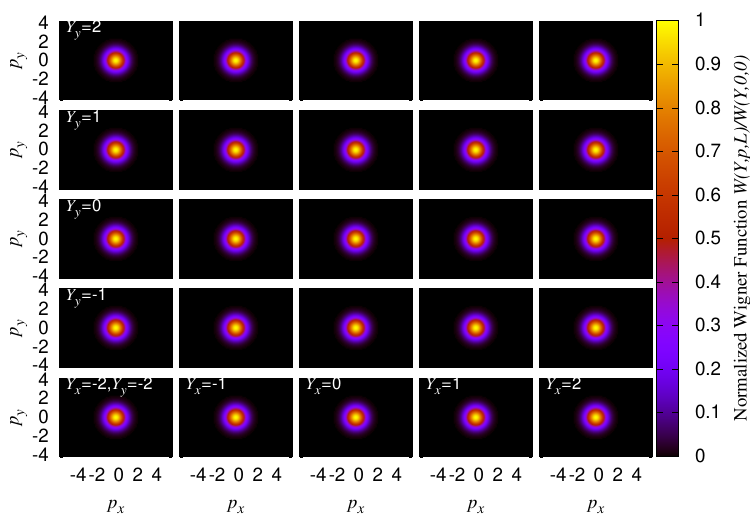}
    \includegraphics[width=0.86\textwidth]{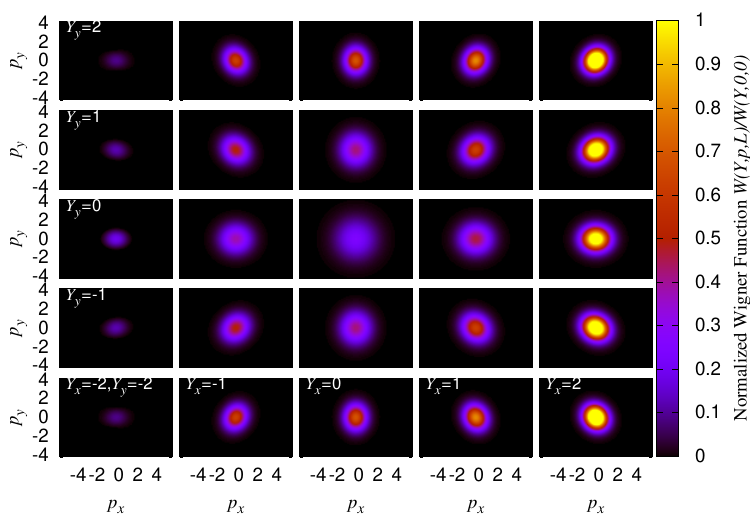}
    \caption{Same plots as in Fig.~\ref{fig:strobo_Flow_constant1}, but including the gradient corrections on the right hand side of Eq.~(\ref{eq:equation_with_flow}).}
    \label{fig:strobo_RHS_Flow_constant1}
\end{figure}

\begin{figure}[]
    \centering
    \includegraphics[width=0.86\textwidth]{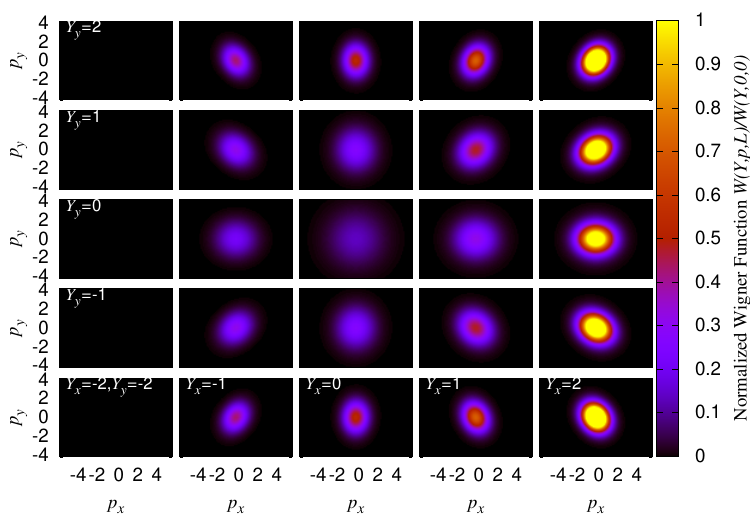}
    \includegraphics[width=0.86\textwidth]{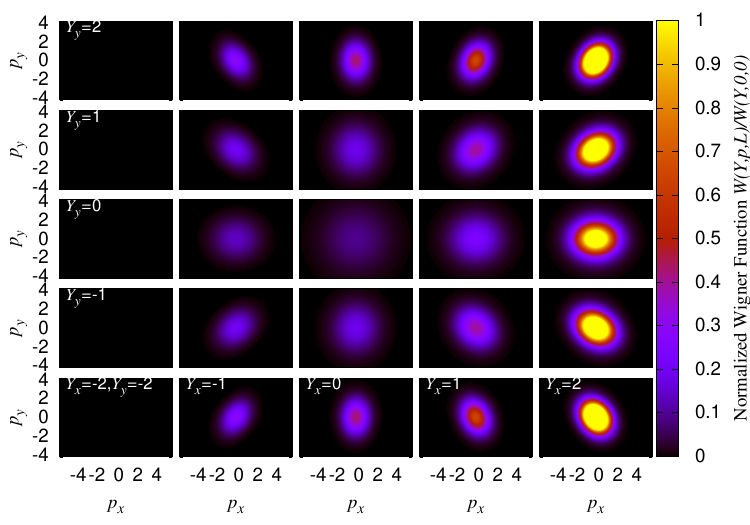}
    \caption{Same results as in Fig.~\ref{fig:strobo_RHS_Flow_constant1}, the upper and lower plots are for $L=2$ and $L=3$, respectively.}
    \label{fig:strobo_RHS_Flow_constant2}
\end{figure}

In contrast, Figs.~\ref{fig:strobo_RHS_Flow_constant1} and \ref{fig:strobo_RHS_Flow_constant2} show the corresponding distributions after evolving the Wigner function using the full Eq.~\eqref{eq:equation_with_flow}. Again, the overall trend of skewing the Wigner distribution along the flow direction with increasing $L$ persists. However, as time progresses, the distribution at a fixed impact parameter also undergoes modifications.
This again highlights that the effects of gradients, beyond those that can be incorporated into $\hat q$, are crucial for properly describing the detailed internal structure of jets.

\subsubsection{Hydrodynamic flow}

\begin{figure}[]
    \centering
    \includegraphics[width=0.49\textwidth]{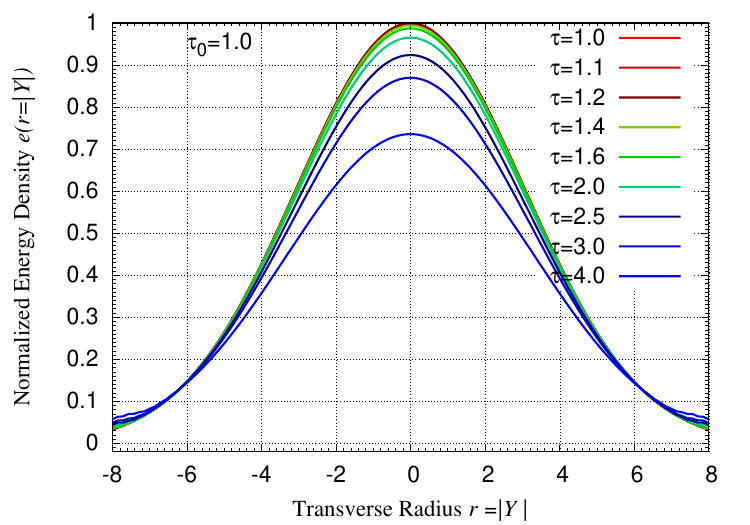}
    \includegraphics[width=0.49\textwidth]{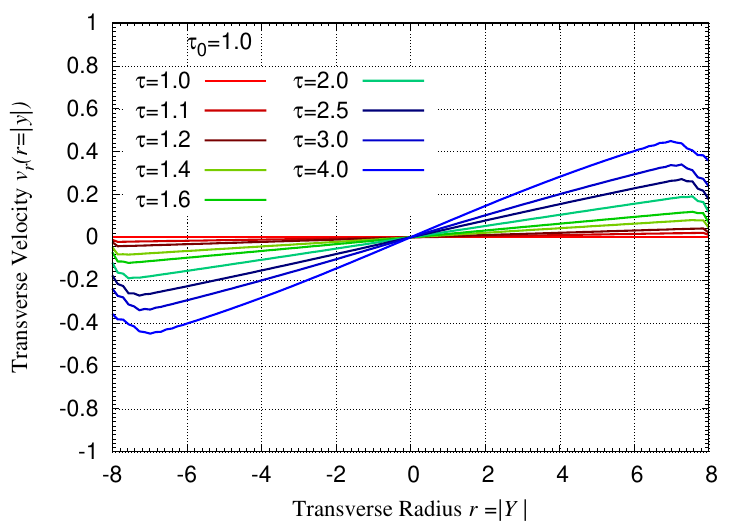}
    \caption{Hydrodynamic evolution of energy distribution $e(\tau,|\Y|)$, and flow velocity $v_r(\tau,|\Y|)$ evaluated at the $x-y$ plane of a Minkowski space $\mathbb{R}^{3,1}$ with boost invariance in the longitudinal direction such that $v_z=0$. 
    The initial profile for the hydrodynamic evolution is isotropic with number density $\rho(\Y)=\rho_0\exp(-\mu_x^2 Y_x^2-\mu_y^2 Y_y^2)$ and $\mu_x=\mu_y=0.2$. The isotropic medium profile enforces an azimuthally symmetric flow generation pattern.}
    \label{fig:Hydro_mu0.2}
\end{figure}

Finally, we study the full transport equation given by Eq.~\eqref{eq:equation_with_flow} in the presence of a hydrodynamic background. Following the above considerations regarding the validity of the evolution equation considering the smallness of gradients, in Fig.~\ref{fig:Hydro_mu0.2} we show the evolution of the energy density and the flow field. Considering the conditions for the validity of the gradient expansion determining Eq.~\eqref{eq:equation_with_flow}, we now use a broader initial distribution compared to the static medium case.

\begin{figure}[]
    \centering
    \includegraphics[width=0.86\textwidth]{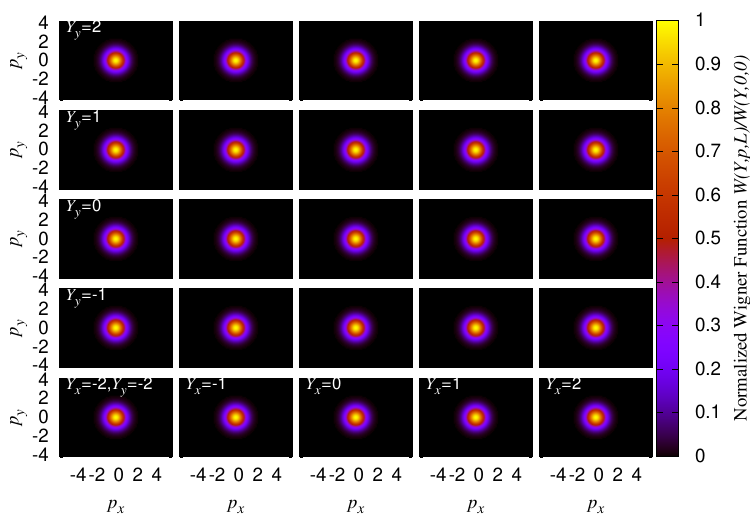}
    \includegraphics[width=0.86\textwidth]{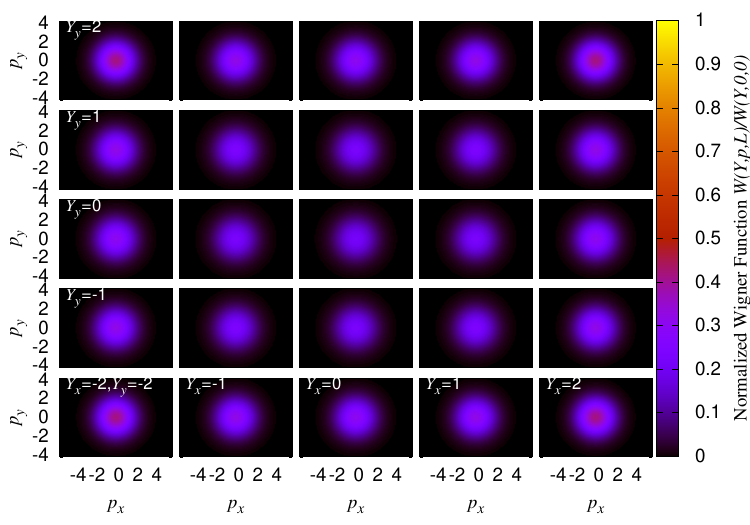}
    \caption{Heatmap of the Wigner function at various positions. We include all gradient effects, the medium expansion, and the matter flow, according to Eq.~(\ref{eq:equation_with_flow}). 
    The flow is assumed to be a hydrodynamic, with the initial condition $\mu_x=\mu_y=0.2$ presented in Fig~\ref{fig:Hydro_mu0.2}. The upper (lower) plot is for $L=0$ ($L=1$).}
    \label{fig:strobo_RHS_Flow_hydro_mu0.2_1}
\end{figure}

\begin{figure}[]
    \centering
    \includegraphics[width=0.86\textwidth]{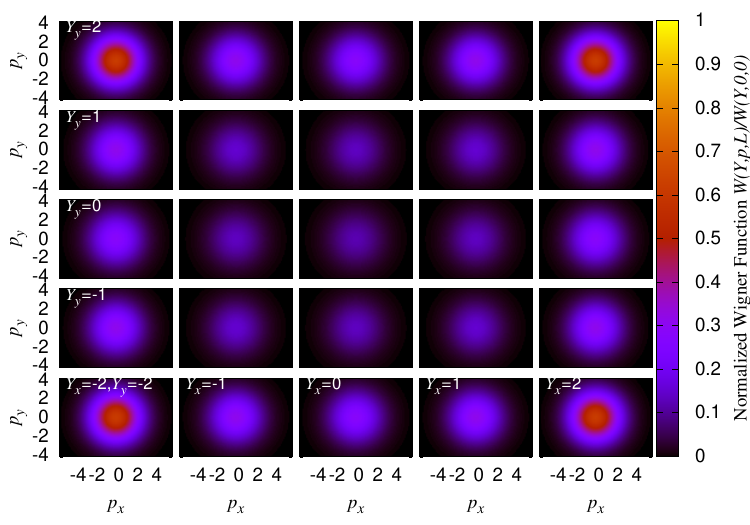}
    \includegraphics[width=0.86\textwidth]{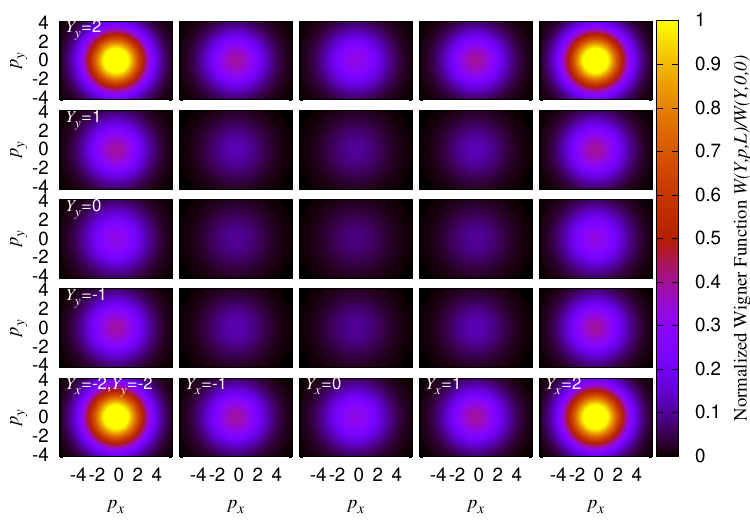}
    \caption{Same results as in Fig.~\ref{fig:strobo_RHS_Flow_hydro_mu0.2_1}, the upper (lower) plot is for $L=2$ ($L=3$).}
    \label{fig:strobo_RHS_Flow_hydro_mu0.2_2}
\end{figure}

\begin{figure}[]
    \centering
    \includegraphics[width=0.86\textwidth]{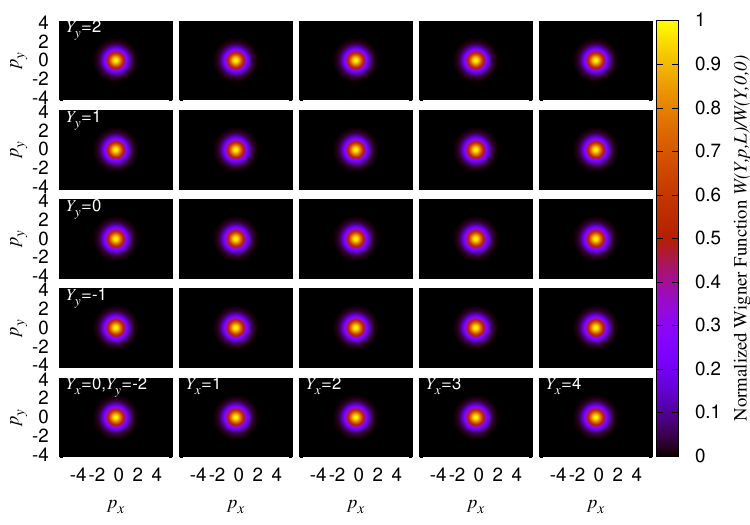}
    \includegraphics[width=0.86\textwidth]{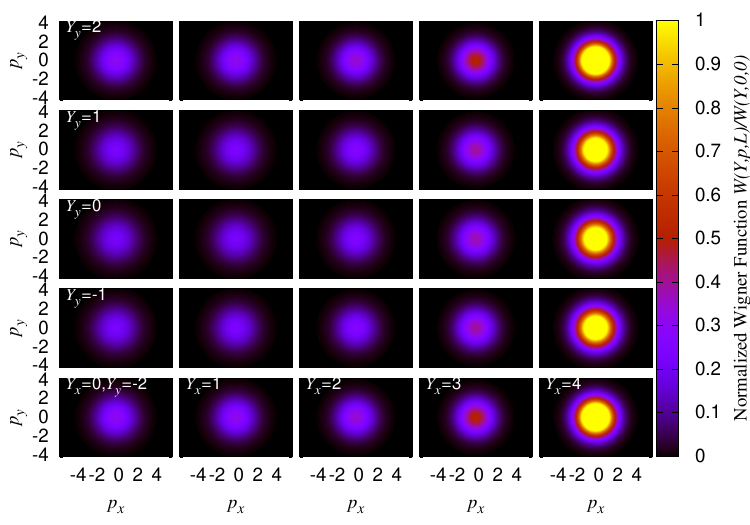}
    \caption{Same results as in Fig.~\ref{fig:strobo_RHS_Flow_hydro_mu0.2_1} but with the jet Wigner function initially centered at $Y_x=1,Y_y=0$. The upper (lower) plot is for $L=0$ ($L=1$).}
    \label{fig:strobo_RHS_Flow_hydro_mu0.2_off_x1.0_1}
\end{figure}

In Figs.~\ref{fig:strobo_RHS_Flow_hydro_mu0.2_1} and \ref{fig:strobo_RHS_Flow_hydro_mu0.2_2}, we again present the spacetime evolution of the Wigner distribution. As with the simpler case of a fixed flow considered above, the overall momentum distribution becomes skewed in the direction of the background flow, i.e. away from the center in this case. The anisotropy of the distribution within each panel, for a fixed impact parameter, is qualitatively much smaller compared to the previous case. This is expected, as the jet direction is directly aligned with the \textit{center} of the flow, making this the most isotropic possible setup. In a more realistic case, where the jet direction and the velocity field are misaligned, additional skewness would be generated. We illustrate such a case in Fig.~\ref{fig:strobo_RHS_Flow_hydro_mu0.2_off_x1.0_1},  where the center of the hard parton Wigner function is set at $(Y_x,Y_y)=(2,0)$.

%%%%%%

\section{Conclusion}\label{sec:conclusion}

In this work, we have presented a numerical study of hard parton transport in an inhomogeneous and/or anisotropic QCD background. To that end, we have numerically solved the transports equations introduced in~\cite{Barata:2022utc}, accounting for medium gradients beyond the classical Boltzmann/AMY framework. As demonstrated through a variety of observables, the inclusion of these corrections is not merely a theoretical concern; it is, in fact, critical to properly describing the evolution of hard probes in anisotropic backgrounds. Moreover, such scenarios are found not only in the QGP phase but also in all stages of heavy-ion collisions.

Translating our findings to more phenomenological and experimental contexts, properly incorporating matter structure effects into the description of jets in off-central high energy heavy-ion collisions and to smaller systems~\cite{Nagle:2018nvi, Arslandok:2023utm,Apolinario:2022vzg} is crucial for making quantitative predictions. Moreover, any full-scale simulation of jet quenching physics that neglects such corrections cannot provide a \textit{bona fide} description of jets. In particular, this means that if one uses simulation codes where the matter structure enters solely through $\hat q$, the description is strictly only accurate at leading gradient order. When flow effects are included, the description becomes more intricate \cite{Kuzmin:2023hko}, and gradients appear to play a crucial role in capturing the full momentum distribution. We note that in small collisional systems, as the ones to be explored in future light ion LHC runs~\cite{Brewer:2021kiv} and in proton-ion events~\cite{Arslandok:2023utm}, the produced hot QCD matter state is anticipated to be far out of equilibrium, dominated by large pressure gradients. Thus, we expect our findings to drive new developments in the description of hard probes in such systems.

In the future, we hope to extend the consideration here to more realistic scenarios, incorporating our results into existing full-scale jet quenching simulation packages. In their current form, implementing the results should be straightforward, though a detailed study of the validity of the approximations used—such as the smallness of gradients—in more complex geometries and backgrounds is warranted. In addition, the same type of corrections need to be incorporated into inelastic processes responsible for the production of induced radiation, see e.g. \cite{Barata:2023zqg} for a recent discussion.

\begin{acknowledgments}
We thank X.-N. Wang for multiple helpful discussions on jet quenching theory. XD is grateful to Lipei Du for valuable discussions on hydrodynamic numerics, and thanks S\"oren Schlichting for helpful comments. XD and AVS acknowledge the support from European Research Council project ERC-2018-ADG-835105 YoctoLHC. XD is also supported by the Spanish Research State Agency under project PID2020-119632GB-I00; by Xunta de Galicia (Centro singular de investigacion de Galicia accreditation 2019-2022), and by European Union ERDF. The work of AVS is supported by Funda\symbol{"00E7}\symbol{"00E3}o para a Ci\symbol{"00EA}ncia e a Tecnologia (FCT) under contract 2022.06565.CEECIND.

\end{acknowledgments}

\bibliographystyle{bibstyle}
\bibliography{refs.bib}

\appendix

\pagebreak

\section{Numerical algorithm of the master equation solver}\label{app:lattice}
In this Appendix, we provide details on the numerical routines used to solve the evolution equation (Eq.~(\ref{eq:numerics_dif_eq})) in the main text.

\subsection{Alternative direction implicit (ADI) algorithm}
The master equation Eq.~(\ref{eq:numerics_dif_eq}) can be decomposed into an expansion term $\pv\cdot\tvec{\na}_\Y/E$ and the diffusion term $D_{ab}\, \partial^2/\partial p_a\partial p_b$.
Both the expansion terms and the off-diagonal terms in diffusion $D_{ab}$ with $a\ne b$ can be easily simulated with the Forward Euler algorithm.
The diagonal terms in the diffusion equation
\begin{align}
\nonumber
\frac{\partial U}{\partial t}=D_{xx}\frac{\partial^2 U}{\partial x^2}
+D_{yy}\frac{\partial^2 U}{\partial y^2}\, ,
\end{align}
can be solved with Forward Euler only if it meets the Courant–Friedrichs–Lewy (CFL) condition $\Delta L < C\Delta x^2$ where the coefficient $C$ can be different at different locations in the medium.
To avoid the complication in choosing the CFL coefficient, an alternative and unconditionally stable algorithm called the alternative direction implicit (ADI) algorithm is used.
The 2D Alternative direction implicit discretization includes two sequential steps
\begin{align}
\nonumber
\frac{U_{i,j}^{t+1/2}-U_{i,j}^{t}}{\Delta t/2}&=D_{xx}\frac{U_{i+1,j}^{t+1/2}-2U_{i,j}^{t+1/2}+U_{i-1,j}^{t+1/2}}{\Delta x^2}
+D_{yy}\frac{U_{i,j+1}^{t}-2U_{i,j}^{t}+U_{i,j-1}^{t}}{\Delta y^2}\, ,\\
\nonumber
\frac{U_{i,j}^{t+1}-U_{i,j}^{t+1/2}}{\Delta t/2}&=D_{xx}\frac{U_{i+1,j}^{t+1/2}-2U_{i,j}^{t+1/2}+U_{i-1,j}^{t+1/2}}{\Delta x^2}
+D_{yy}\frac{U_{i,j+1}^{t+1}-2U_{i,j}^{t+1}+U_{i,j-1}^{t+1}}{\Delta y^2}\, .
\end{align}
Denoting $\alpha=\frac{D_{xx}\Delta t}{\Delta x^2}$, $\beta=\frac{D_{yy}\Delta t}{\Delta y^2}$, the matrix equation we want to solve is 
\begin{align}
\nonumber
2(1+\alpha)U_{i,j}^{t+1/2}-\alpha U_{i+1,j}^{t+1/2}-\alpha U_{i-1,j}^{t+1/2}&=2(1-\beta)U_{i,j}^{t}+\beta U_{i,j+1}^{t}+\beta U_{i,j-1}^{t}=B^{t}\, ,\\
\nonumber
2(1+\beta)U_{i,j}^{t+1}-\beta U_{i,j+1}^{t+1}-\beta U_{i,j-1}^{t+1}&=2(1-\alpha) U_{i,j}^{t+1/2}+\alpha U_{i+1,j}^{t+1/2}+\alpha U_{i-1,j}^{t+1/2}=B^{t+1/2}\, .
\nonumber
\end{align}
The above equations can be presented in the following tridiagonal matrix form and it is easy to solve for each direction
\begin{align}
\nonumber
\begin{pmatrix}
2(1+\alpha) & -\alpha & 0 \\
-\alpha & 2(1+\alpha) & -\alpha  \\
0 & -\alpha & 2(1+\alpha) \\
\end{pmatrix}
\begin{pmatrix} 
U_{i-1,j}^{t+1/2} \\
U_{i,j}^{t+1/2} \\
U_{i+1,j}^{t+1/2} \\
\end{pmatrix}&=B^{t} \, ,\\
\nonumber
\begin{pmatrix} 
2(1+\beta) & -\beta & 0 \\
-\beta & 2(1+\beta) & -\beta \\
0 & -\beta & 2(1+\beta) \\
\end{pmatrix}
\begin{pmatrix} 
U_{i,j-1}^{t+1} \\
U_{i,j}^{t+1} \\
U_{i,j+1}^{t+1} \\
\end{pmatrix}&=B^{t+1/2} \, .
\end{align}

\subsection{Tridiagonal matrix algorithm}
The above ADI requires a solution to the tridiagonal matrix equation $AU^{t+1}=B^{t}$ which can be solved by the LU~decomposition $A=\bar{L}\bar{U}$. We solve $\bar{L}Y=B$ then $\bar{U}U=Y$
\begin{eqnarray}
\nonumber
\begin{pmatrix} 
b_{1} & c_{1} & 0 & 0 & \cdots & 0 \\
a_{2} & b_{2} & c_{2} & 0 & \cdots & 0 \\
\vdots & \vdots & \vdots & \vdots & \ddots & \vdots \\
0 & \cdots & 0 & a_{n-1} & b_{n-1} & c_{n-1} \\ 
0 & \cdots & 0 & 0 & a_{n} & b_{n} \\ 
\end{pmatrix}
=
\begin{pmatrix} 
1 & 0 & 0 & 0 & \cdots & 0 \\
l_{2} & 1 & 0 & 0 & \cdots & 0 \\
\vdots & \vdots & \vdots & \vdots & \ddots & \vdots \\
0 & \cdots & 0 & l_{n-1} & 1 & 0 \\ 
0 & \cdots & 0 & 0 & l_{n} & 1 \\ 
\end{pmatrix}
\begin{pmatrix} 
v_{1} & c_{1} & 0 & 0 & \cdots & 0 \\
0 & v_{2} & c_{2} & 0 & \cdots & 0 \\
\vdots & \vdots & \vdots & \vdots & \ddots & \vdots \\
0 & \cdots & 0 & 0 & v_{n-1} & c_{n-1} \\ 
0 & \cdots & 0 & 0 & 0 & v_{n} \\ 
\end{pmatrix} \, ,
\end{eqnarray}
The $\bar{L},\bar{U}$ can be calculated as
\begin{eqnarray}
\nonumber
    b_{1}=v_{1} &\to& v_{1}=b_{1} \, ,\\
    \nonumber
    a_{k}=l_{k}v_{k-1} &\to& l_{k}=a_{k}/v_{k-1},~~~k=2,...,n\, ,\\
    \nonumber
    b_{k}=l_{k}c_{k-1}+v_{k} &\to& v_{k}=b_{k}-l_{k}c_{k-1},~~~k=2,...,n \, .
\end{eqnarray}
To solve $\bar{L}Y=B$, we use
\begin{eqnarray}
\nonumber
    Y_{1}=B_{1} &\to& Y_{1}=B_{1}\, , \\
    \nonumber
    l_{k}Y_{k-1}+Y_{k}=B_{k} &\to& Y_{k}=B_{k}-l_{k}Y_{k-1},~~~k=2,...,n \, .
\end{eqnarray}
To solve $\bar{U}U=Y$, we have
\begin{eqnarray}
\nonumber
    v_{n}U_{n}=Y_{n} &\to& U_{n}=Y_{n}/v_{n}\, ,\\
    \nonumber
    v_{k}U_{k}+c_{k}U_{k+1}=Y_{k} &\to& U_{k}=(Y_{k}-c_{k}U_{k+1})/v_{k},~~~k=n-1,...,1\, , 
\end{eqnarray}
Thus, we find $U^{t+1}=\bar{U}^{-1}L^{-1}B$.

\pagebreak
\section{Numerical algorithm of the hydrodynamic equation solver}
~\label{app:hydro}
In this section, we present the numerical algorithm that we have used in solving the relativistic ideal hydrodynamic (Euler fluid) equation in 2+1 dimension, in both Minkowski and Milne coordinates.
In Milne coordinates, there are finite Christoffel connections and the additional terms associated are labeled in blue for the benefit of audiences less familiar with hydrodynamic theory.

\subsection{Hydrodynamic equations}
The hydrodynamic equations are
\begin{align}
\nonumber
D_{\mu}T^{\mu\nu}&=\partial_{\mu}T^{\mu\nu}+\textcolor{blue}{\Gamma^{\mu}_{\mu\lambda}T^{\lambda\nu}+\Gamma^{\nu}_{\mu\lambda}T^{\mu\lambda}}=0\, ,\nn\nonumber
D_{\mu}J^{\mu}&=\partial_{\mu} J^{\mu}+\textcolor{blue}{\Gamma^{\mu}_{\mu\lambda}J^{\lambda}}=0 \, .
\end{align}
We neglect conserved currents (typically charge) in our discussion, $J^{\mu}=0$, and focus on the evolution of the energy-momentum tensor $T^{\mu\nu}$.
In Milne coordinates, the nontrivial Christoffel connections are $\Gamma^{\tau}_{\eta\eta}=\tau,~ \Gamma^{\eta}_{\eta\tau}=\Gamma^{\eta}_{\tau\eta}=1/\tau$, which lead to
\begin{align}
\nonumber
    \partial_{\mu}T^{\mu\tau}+\textcolor{blue}{\frac{1}{\tau}T^{\tau\tau}}+\textcolor{blue}{\tau T^{\eta\eta}}=0 \, ,~~~
    \partial_{\mu}T^{\mu i}+\textcolor{blue}{\frac{1}{\tau}T^{\tau i}}=0 \, ,~~~
    \partial_{\mu}T^{\mu\eta}+\textcolor{blue}{\frac{3}{\tau}T^{\tau\eta}}=0 \, .
\end{align}
Considering the velocity $v^i=u^i/u^{\tau}$ and the energy-momentum tensor, one has
\begin{align}
\nonumber
    T^{\mu\nu}&=(e+p)u^{\mu}u^{\nu}-pg^{\mu\nu},\\
    \nonumber
    T^{\tau\tau}&=(e+p)(u^{\tau})^2-p,~~~
    T^{\tau i}=(e+p)u^{\tau}u^{i},~~~
    T^{\tau\eta}=(e+p)u^{\tau}u^{\eta},\\
    \nonumber
    T^{ii}&=(e+p)(u^{i})^2+p,~~~T^{ij}=(e+p)u^{i}u^{j},~~~
    T^{i\eta}=(e+p)u^{i}u^{\eta},~~~
    T^{\eta\eta}=(e+p)u^{\eta}u^{\eta}+\frac{p}{\textcolor{blue}{\tau^2}}
\end{align}
such that $T^{\tau i}=v_i T^{\tau\tau}+v_ip$, $T^{ii}=v^{i}T^{\tau i}+p$, $T^{ji}=v^{j}T^{\tau i}$, $T^{i\eta}=v^{i}T^{\tau\eta}$, $T^{\eta\eta}=v^{\eta}T^{\tau\eta}+p/\textcolor{blue}{\tau^2}$, $T^{\tau\eta}=v_{\eta}T^{\tau\tau}+v_{\eta}p$. 
That means we can rewrite the ideal hydrodynamic equations into~\cite{Bazow:2016yra} 
\begin{align}
    \nonumber
    \partial_{\tau}T^{\tau\tau}
    &=
    -\partial_{x}(v_xT^{\tau\tau})
    -\partial_{y}(v_yT^{\tau\tau})
    -\partial_{\eta}(v_{\eta}T^{\tau\tau})
    -\partial_x(v_x p)
    -\partial_y(v_y p)
    -\partial_{\eta}(v_{\eta} p)
    -\textcolor{blue}{\frac{1}{\tau}T^{\tau\tau}-\tau T^{\eta\eta}}\\
    \nonumber
    \partial_{\tau}T^{\tau x}
    &=
    -\partial_{x}(v_xT^{\tau x})
    -\partial_{y}(v_yT^{\tau x})
    -\partial_{\eta}(v_{\eta}T^{\tau x})
    -\partial_x(p)
    -\textcolor{blue}{\frac{1}{\tau}T^{\tau x}}\\
    \nonumber
    \partial_{\tau}T^{\tau y}
    &=
    -\partial_{x}(v_xT^{\tau y})
    -\partial_{y}(v_yT^{\tau y})
    -\partial_{\eta}(v_{\eta}T^{\tau y})
    -\partial_y(p)
    -\textcolor{blue}{\frac{1}{\tau}T^{\tau y}}\\ \nonumber
    \partial_{\tau}T^{\tau \eta}
    &=
    -\partial_{x}(v_xT^{\tau \eta})
    -\partial_{y}(v_yT^{\tau \eta})
    -\partial_{\eta}(v_{\eta}T^{\tau \eta})
    -\frac{1}{\textcolor{blue}{\tau^2}}\partial_{\eta}(p)
    -\textcolor{blue}{\frac{3}{\tau}T^{\tau \eta}}
\end{align}
In what we are interested in, we consider a longitudinally boost-invariant medium such that we can practically set $v_{\eta}=0$ in the numerical simulations.

\subsection{Kurganov-Tadmor (KT) scheme}
We use the standard Kurganov-Tadmor (KT) scheme~\cite{Kurganov:1999} to solve the hydrodynamic equation, although a simple finite difference central scheme is already good enough to solve the ideal fluid. 
We follow the specific implementation discussed in the literature~\cite{Bazow:2016yra} and one may also find more details including viscous hydrodynamic simulations there. 
Although the algorithm is well documented in the literature, we still summarize it below to make the paper self-contained, and more clear.

Denoting the vector of tensors $\mathbf{T}=(T^{\tau\tau},T^{\tau x},T^{\tau y},T^{\tau \eta})$, we need to solve the set of hydrodynamic equations as follows
\begin{align}
\nonumber
   \frac{\partial \mathbf{T}_{i,j,k}}{\partial \tau}&=-\frac{H_{i+1/2,j,k}-H_{i-1/2,j,k}}{\Delta x}
   -\frac{H_{i,j+1/2,k}-H_{i,j-1/2,k}}{\Delta y} 
   -\frac{H_{i,j,k+1/2}-H_{i,j,k-1/2}}{\Delta \eta} +\mathbf{J}^{\tau}_{i,j,k} \, .
\end{align}
It contains a conservative form of energy-momentum tensor evolution, with contributiosn from derivative of tensors $\nabla (\vec{v}\mathbf{T})$ and contribution from sources $\mathbf{J}$.
The source terms contain all contributions not explicitly on the tensor $\mathbf{T}$, for example the derivative of pressure $\partial_{\eta}(v_{\eta}p)$, etc. .
The function $H$ should be evaluated from the staggered tensor and flow velocity
\begin{align}
    \nonumber
    H_{i+1/2,j,k}&=\frac{1}{2}[(v_{x})_{i+1/2,j,k}^{L}\mathbf{T}_{i+1/2,j,k}^{L}+(v_{x})_{i+1/2,j,k}^{R}\mathbf{T}_{i+1/2,j,k}^{R}-a_{i+1/2,j,k}(\mathbf{T}_{i+1/2,j,k}^{R}-\mathbf{T}_{i+1/2,j,k}^{L})]\\
    \nonumber
    H_{i-1/2,j,k}&=\frac{1}{2}[(v_{x})_{i-1/2,j,k}^{L}\mathbf{T}_{i-1/2,j,k}^{L}+(v_{x})_{i-1/2,j,k}^{L}\mathbf{T}_{i-1/2,j,k}^{R}-a_{i-1/2,j,k}(\mathbf{T}_{i-1/2,j,k}^{R}-\mathbf{T}_{i-1/2,j,k}^{L})]\\
    \nonumber
    H_{i,j+1/2,k}&=\frac{1}{2}[(v_{y})_{i,j+1/2,k}^{L}\mathbf{T}_{i,j+1/2,k}^{L}+(v_{y})_{i,j+1/2,k}^{R}\mathbf{T}_{i,j+1/2,k}^{R}-a_{i,j+1/2,k}(\mathbf{T}_{i,j+1/2,k}^{R}-\mathbf{T}_{i,j+1/2,k}^{L})]\\
    \nonumber
    H_{i,j-1/2,k}&=\frac{1}{2}[(v_{y})_{i,j-1/2,k}^{L}\mathbf{T}_{i,j-1/2,k}^{L}+(v_{y})_{i,j+1/2,k}^{R}\mathbf{T}_{i,j-1/2,k}^{R}-a_{i,j-1/2,k}(\mathbf{T}_{i,j-1/2,k}^{R}-\mathbf{T}_{i,j-1/2,k}^{L})]\\
    \nonumber
    H_{i,j,k+1/2}&=0\\
    \nonumber
    H_{i,j,k-1/2}&=0
\end{align}
The $H$ in the longitudinal direction is trivial due to the fact that $|v_z|=0$. 
The propagation speed $a$ at each stagger grid reads
\begin{align}
    \nonumber
    a_{i\pm 1/2,j,k}^{n}&=\max\{|(v_{x})_{i\pm 1/2,j,k}^{L}|,|(v_{x})_{i\pm 1/2,j,k}^{R}|\}\\
    \nonumber
    a_{i,j\pm 1/2,k}^{n}&=\max\{|(v_{y})_{i,j\pm 1/2,k}^{L}|,|(v_{y})_{i,j\pm 1/2,k}^{R}|\}\\ \nonumber
    a_{i,j,k\pm 1/2}&=|v_{\eta}|=0
\end{align}
The piecewise interpolant of the tensor (take $x$ direction for example) $\mathbf{T}_{i\pm1/2,j,k}^{L,R}$, at the staggered cell can be calculated as
\begin{align}
\nonumber
    \mathbf{T}_{i+1/2,j,k}^{L}&=\bar{\mathbf{T}}_{i,j,k}+\frac{\Delta x}{2}(\mathbf{T}_x)_{i,j,k},~~~~~~~~~~~
    \mathbf{T}_{i+1/2,j,k}^{R}=\bar{\mathbf{T}}_{i+1,j,k}-\frac{\Delta x}{2}(\mathbf{T}_x)_{i+1,j,k},\\
    \nonumber
    \mathbf{T}_{i-1/2,j,k}^{L}&=\bar{\mathbf{T}}_{i-1,j,k}+\frac{\Delta x}{2}(\mathbf{T}_x)_{i-1,j,k},~~~~~~
    \mathbf{T}_{i-1/2,j,k}^{R}=\bar{\mathbf{T}}_{i,j,k}-\frac{\Delta x}{2}(\mathbf{T}_x)_{i,j,k},\\
    \nonumber
    \mathbf{T}_{i,j+1/2,k}^{L}&=\bar{\mathbf{T}}_{i,j,k}+\frac{\Delta y}{2}(\mathbf{T}_y)_{i,j,k},~~~~~~~~~~~
    \mathbf{T}_{i,j+1/2,k}^{R}=\bar{\mathbf{T}}_{i,j+1,k}-\frac{\Delta y}{2}(\mathbf{T}_y)_{i,j+1,k}\\
    \nonumber
    \mathbf{T}_{i,j-1/2,k}^{L}&=\bar{\mathbf{T}}_{i,j-1,k}+\frac{\Delta y}{2}(\mathbf{T}_y)_{i,j-1,k},~~~~~~
    \mathbf{T}_{i,j-1/2,k}^{R}=\bar{\mathbf{T}}_{i,j,k}-\frac{\Delta y}{2}(\mathbf{T}_y)_{i,j,k}
\end{align}
The sliding average $\bar{\mathbf{T}}$ that smooths any shock wave is defined as
\begin{align}
\nonumber
    \bar{\mathbf{T}}_{i,j,k}&=\frac{1}{\Delta x\Delta y\Delta z}\int_{x_i-\Delta x/2}^{x_i+\Delta x/2}\int_{y_i-\Delta y/2}^{y_i+\Delta y/2}\int_{z_i-\Delta \eta/2}^{z_i+\Delta \eta/2}\mathbf{T}(x,y,\eta)dxdydz
\end{align}
With linear interpolation for $(x,y,\eta)\in [x_{i},x_{i+1}]\times[y_{i},y_{i+1}]\times[\eta_{i},\eta_{i+1}]$, the function $\mathbf{T}$ is evaluated as
\begin{align}
    \nonumber
    \mathbf{T}(x,y,\eta)&=w_L(x)w_L(y)w_L(\eta)\mathbf{T}_{i,j,k}+w_L(x)w_L(y)w_R(\eta)\mathbf{T}_{i,j,k+1}\\
    \nonumber&+w_L(x)w_R(y)w_L(\eta)\mathbf{T}_{i,j+1,k}+w_L(x)w_R(y)w_R(\eta)\mathbf{T}_{i,j+1,k+1}\\
    \nonumber&+w_R(x)w_L(y)w_L(\eta)\mathbf{T}_{i+1,j,k}+w_R(x)w_L(y)w_R(\eta)\mathbf{T}_{i+1,j,k+1}\\ \nonumber
    &+w_R(x)w_R(y)w_L(\eta)\mathbf{T}_{i+1,j+1,k}+w_R(x)w_R(y)w_R(\eta)\mathbf{T}_{i+1,j+1,k+1}
\end{align}
where the weight functions are defined as
\begin{align}
\nonumber
    w_L(x)=\frac{x_{i+1}-x}{x_{i+1}-x_{i}},~~~~~~
    w_R(x)=\frac{x-x_{i}}{x_{i+1}-x_{i}}
\end{align}
The integration for the sliding average can be written in an algebraic expression, although it is pretty lengthy.
The minmod limiter $\mathbf{T}_x$ is defined as
\begin{align}
    \nonumber
    (\mathbf{T}_x)_{i,j,k}&={\rm minmod}\bigg(\frac{\mathbf{T}_{i,j,k}-\mathbf{T}_{i-1,j,k}}{\Delta x},\frac{\mathbf{T}_{i+1,j,k}-\mathbf{T}_{i,j,k}}{\Delta x}\bigg),\\ \nonumber
    {\rm minmod}(a,b)&=\frac{1}{2}[{\rm sgn}(a)+{\rm sgn}(b)]\cdot{\rm min}(|a|,|b|).
\end{align}
Although the conservative terms are evaluated from staggered cells, the source terms can be easily evaluated from a simple mid-scheme finite difference method
 for example
\begin{align}
\nonumber
    \partial_{\eta}(v_{\eta}p)_{i,j,k}=\frac{(v_{\eta}p)_{i,j,k+1}-(v_{\eta}p)_{i,j,k-1}}{2\Delta \eta}.
\end{align}
In the above discussions of the finite-difference method and staggered cells, we implement ``ghost cells" at the boundary with equal value to the edge of the physical cells.
The time evolution is performed with a 2nd-order Runge-Kutta method
\begin{align}
    \nonumber
    \mathbf{T}^{n+1/2}&=\mathbf{T}^{n}+\Delta\tau \mathbf{C}[\mathbf{T}^{n}]\\ \nonumber
    \mathbf{T}^{n+1}&=\frac{1}{2}\mathbf{T}^{n}+\frac{1}{2}(\mathbf{T}^{n+1/2}+\Delta\tau \mathbf{C}[\mathbf{T}^{n+1/2}])
\end{align}
Finally, after the evolution, the pressure and the flow velocity can be calculated algebraically with a conformal equation of state (EoS) $e=3p$
\begin{align}
\nonumber
    p=\frac{\sqrt{4(T^{\tau\tau})^2-3M^2}-T^{\tau\tau}}{3},~~~
    u^{\tau}=\sqrt{\frac{T^{\tau\tau}+p}{e+p}},~~~
    u^{i}
    =\frac{T^{\tau i}}{(e+p)u^{\tau}},~~~
    v_{i}=\frac{u^{i}}{u^{\tau}}
\end{align}
with $M^2=(T^{\tau x})^2+(T^{\tau y})^2+\textcolor{blue}{\tau^2}(T^{\tau \eta})^2$.

We perform the simulation With lattice $N_x \times N_y \times N_{\eta}=111\times 111 \times 1$ on space $(x,y,\eta)\in[-8,8]\times[-8,8]\times 0$.
Since we have equations roughly $\partial_{\tau}\mathbf{T}\simeq -\nabla v\mathbf{T}$, the CFL condition for a stable hydrodynamic simulation in our case is quite simple $\Delta\tau\lesssim\Delta x/v$. Since the velocity must be smaller than unity $v<1$, one can just choose $\Delta\tau\lesssim\Delta x$. In practice, we choose a $\Delta\tau$ that is $10$ times smaller than this bound.

\subsection{Hydrodynamic simulation test with Gubser flow}
To validate the hydrodynamic simulation, the standard benchmark test~\cite{Nopoush:2014qba,Pang:2014ipa,Du:2019obx} is to compare that with the Gubser flow~\cite{Gubser:2010ze,Gubser:2010ui} which turns out to have an analytical solution. 
The discussions on Gubser flow can be found in the literature, and we briefly summarize the key aspects of Gubser flow in this section to make the paper self-contained.

The 4-velocity flow profile in the cylindrical Milne coordinates can be written as $\tilde{u}^{\mu}=(\tilde{u}^{\tau},\tilde{u}^{r},\tilde{u}^{\phi},\tilde{u}^{\eta})$, and the components are
\begin{align}
    \nonumber
    \tilde{u}^{\tau}&=\cosh(\eta)\gamma-\sinh(\eta)\gamma v_z,~~~~~~~~~~~~~
    \tilde{u}^{r}=\cos(\phi)\gamma v_x+\sin(\phi)\gamma v_y,\\
    \nonumber
    \tilde{u}^{\phi}&=-\frac{1}{r}(\sin(\phi)\gamma v_x-\cos(\phi)\gamma v_y),~~~~~~
    \nonumber
    \tilde{u}^{\eta}=-\frac{1}{\tau}(\sinh(\eta)\gamma-\cosh(\eta)\gamma v_z) \, .
\end{align}

The Gubser flow assumes boost invariance in the longitudinal direction such that $\tilde{u}^{\eta}=0$ and azimuthal symmetry in the transverse plane such that $\tilde{u}^{\phi}=0$, but a finite transverse flow $\tilde{u}^{r}\ne 0$. 
Since $\tilde{u}^{\eta}=0$, one immediately realizes that $v_z=\tanh(\eta)=z/t$ and thus $\tilde{u}^{\tau}=\gamma/\cosh(\eta)$. Since $\tilde{u}^{\phi}=0$, one has that $v_y/v_x=\tan(\phi)$. Assuming the transverse velocity to be $v_{r}=\sqrt{v_x^2+v_y^2}$, one has $\tilde{u}^{r}=\gamma v_{r}$. This leads to the fact that Gubser flow satisfies $\tilde{u}_{\rm Gubser}=(\gamma/\cosh(\eta),\gamma v_{r},0,0)$. Since $\tilde{u}^{\mu}\tilde{u}_{\mu}=1$, one has $\gamma^2(1/\cosh^2(\eta)-v_{r}^2)=1$. 
Indeed, the normalization condition for the 4-velocity leads to a form $\gamma/\cosh(\eta)=\cosh(\kappa)$ and $\gamma v_{r}=\sinh(\kappa)$. Together with $v_x=v_{r}\cosh(\phi)$ and $v_y=v_{r}\sin(\phi)$ we have that,
in Minkowski and in Milne coordinates, the Gubser flow is
\begin{align}
\nonumber
{\rm Minkowski:}~u^{\mu}_{\rm Gubser}&=(\cosh(\kappa)\cosh(\eta),\sinh(\kappa)\cos(\phi),\sinh(\kappa)\sin(\phi),\cosh(\kappa)\sinh(\eta)) \, ,\\ \nonumber
{\rm Cylindrical~Milne:}~\tilde{u}^{\mu}_{\rm Gubser}&=(\tilde{u}^{\tau},\tilde{u}^{r},\tilde{u}^{\phi},\tilde{u}^{\eta})
=(\cosh(\kappa),\sinh(\kappa),0,0) \, .
\end{align}
Furthermore, Gubser assumes~\cite{Gubser:2010ze} that 
the velocity profile function/transverse rapidity $\kappa$ with a scaling parameter $q$ that controls transverse expansion rate
\begin{align}
\nonumber
    \tanh(\kappa)&=\frac{2q^2\tau r}{q^2(\tau^2+r^2)+1} \, .
\end{align}
The analytical solution of energy density and velocity profiles can be obtained easily in de Sitter space $dS_3\times\mathbb{R}$. 
In Gubser's paper, it is formulated as a Weyl-rescaled coordinates
\begin{align}
    \nonumber
    \tilde{x}^{\mu}&=(\rho,\theta,\phi,\eta)=
    \bigg(\sinh^{-1}\bigg(\frac{q^2\tau^2-q^2r^2-1}{2q\tau}\bigg),\tan^{-1}\bigg(\frac{2qr}{q^2\tau^2-q^2r^2+1}\bigg),\phi,\eta\bigg) \, ,\\ \nonumber
    ds^2/\tau^2=d\tilde{s}^2&=
    d\rho^2-\cosh^2(\rho)(d\theta^2+\sin^2(\theta)d\phi^2)-d\eta^2 \, .
\end{align}
The metric explicitly reads 
\begin{align}
\nonumber
\tilde{g}_{\mu\nu}&={\rm diag}(+1,-\cosh^2(\rho),-\cosh^2(\rho)\sin^2(\theta),-1),\\~~~~ 
\nonumber
\tilde{g}^{\mu\nu}&={\rm diag}(+1,-\frac{1}{\cosh^2(\rho)},-\frac{1}{\cosh^2(\rho)\sin^2(\theta)},-1) \, .
\end{align}
This leads to nontrivial Christoffel connections
\begin{align}
\nonumber
\Gamma^{\rho}_{\theta\theta}
&=-\frac{1}{2}g^{\rho\rho}\bigg(\frac{\partial g_{\theta\theta}}{\partial x^{\rho}}\bigg)
=\cosh(\rho)\sinh(\rho),~~~~~~
\Gamma^{\rho}_{\phi\phi}
=-\frac{1}{2}g^{\rho\rho}\bigg(\frac{\partial g_{\phi\phi}}{\partial x^{\rho}}\bigg)
=\cosh(\rho)\sinh(\rho)\sin^2(\theta),\\
%%%%%
\nonumber
\Gamma^{\theta}_{\phi\phi}
&=-\frac{1}{2}g^{\theta\theta}\bigg(\frac{\partial g_{\phi\phi}}{\partial x^{\theta}}\bigg)
=-\cos(\theta)\sin(\theta),~~~~~~~
\Gamma^{\phi}_{\phi\theta}
=\frac{1}{2}g^{\phi\phi}\bigg(\frac{\partial g_{\phi\phi}}{\partial x^{\theta}}\bigg)
=\frac{1}{\tan(\theta)} \, , \\  \nonumber
%%%%%
\Gamma^{\theta}_{\theta\rho}
&=\frac{1}{2}g^{\theta\theta}\bigg(\frac{\partial g_{\theta\theta}}{\partial x^{\rho}}\bigg)
=\tanh(\rho),~~~~~~~~~~~~~~~~~~
\Gamma^{\phi}_{\phi\rho}
=\frac{1}{2}g^{\phi\phi}\bigg(\frac{\partial g_{\phi\phi}}{\partial x^{\rho}}\bigg)
=\tanh(\rho) \, .
\end{align}
In the de~Sitter coordinates, Gubser further assumes~\cite{Gubser:2010ui} the flow profile to be $\tilde{u}^{\mu}_{\rm Gubser}=(\tilde{u}^{\rho},\tilde{u}^{\theta},\tilde{u}^{\phi},\tilde{u}^{\eta})
=(1,0,0,0)$. 
Then one can write down the nontrivial terms of energy-momentum tensor
\begin{align}
\nonumber
\tilde{T}^{\rho\rho}_{\rm Gubser}&=e,~~~~~~
\tilde{T}^{\theta\theta}_{\rm Gubser}=\frac{p}{\cosh^2(\rho)},~~~~~~
\tilde{T}^{\phi\phi}_{\rm Gubser}=\frac{p}{\cosh^2(\rho)\sin^2(\theta)},~~~~~~
\tilde{T}^{\eta\eta}_{\rm Gubser}=p \, .
\end{align}
Eventually, the temporal component of the hydrodynamic equation reads
\begin{align}
\nonumber
D_{\mu}T^{\mu\rho}&=\partial_{\rho}T^{\rho\rho}+\Gamma^{\theta}_{\theta\rho}T^{\rho\rho}+\Gamma^{\phi}_{\phi\rho}T^{\rho\rho}+\Gamma^{\rho}_{\theta\theta}T^{\theta\theta}+\Gamma^{\rho}_{\phi\phi}T^{\phi\phi}
=\partial_{\rho} e+ 2e\tanh(\rho)+2p\tanh(\rho)
=0 \, .
\end{align}
With conformal EoS $e=3p$, we arrive at an analytical solution for the energy density
\begin{align}
\nonumber
    \partial_{\rho}\bigg(e\cosh^{8/3}(\rho)\bigg)=0,~~~~~~
    {\rm solution:}~e(\rho)=\bigg(\frac{\cosh(\rho_0)}{\cosh(\rho)}\bigg)^{8/3}e(\rho_0) \, .
\end{align}
Replacing the time in de Sitter space $\rho$ with time $\tau$ and transverse radius $r$ in Milne space, one obtains
\begin{align}
\nonumber
    e(\rho)\simeq\frac{(2q\tau)^{8/3}}{[q^4(\tau^2-r^2)^2+2q^2(\tau^2+r^2)+1]^{4/3}} \, .
\end{align}
The length and time in $\mathbb{R}^{3,1}$ should be rescaled by $\tau$ to $dS_3\times\mathbb{R}$ thus 
to obtain the final solution in $\mathbb{R}^{3,1}$, the length and time should have an inversed Weyl rescaling by $\tau$ from de Sitter space $dS_3\times\mathbb{R}$~\cite{Gubser:2010ui}. Eventually, we have 
\begin{align}
\nonumber
    e_{\rm Gubser}(\tau,r)&\simeq\frac{1}{\tau^4}\frac{(2q\tau)^{8/3}}{[q^4(\tau^2-r^2)^2+2q^2(\tau^2+r^2)+1]^{4/3}} \, ,\\ \nonumber
    v_{r,\rm Gubser}(\tau,r)&=\frac{u^{r}}{u^{\tau}}=\tanh(\kappa)=\frac{2q^2\tau r}{q^2(\tau^2+r^2)+1} \, .
\end{align}

\begin{figure}[h!]
    \centering
    \includegraphics[width=0.495\textwidth]{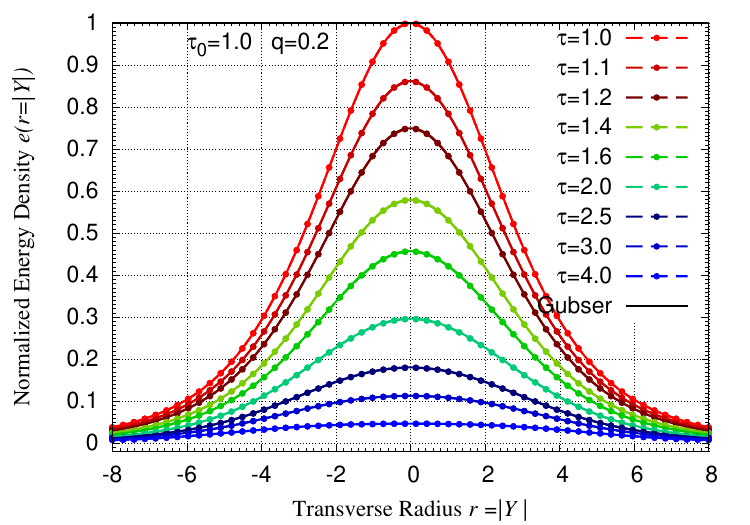}
    \includegraphics[width=0.495\textwidth]{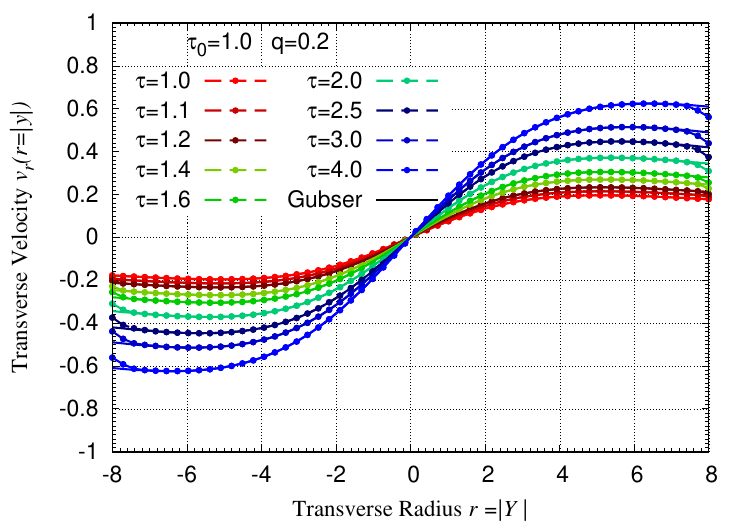}
     \includegraphics[width=0.495\textwidth]{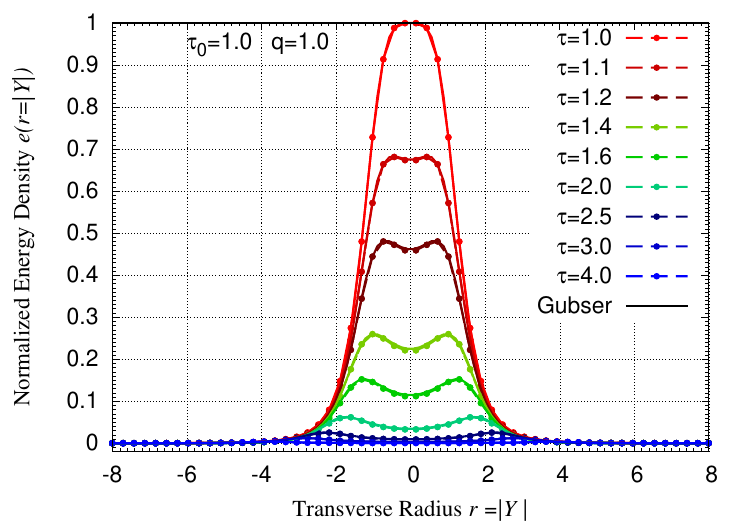}
    \includegraphics[width=0.495\textwidth]{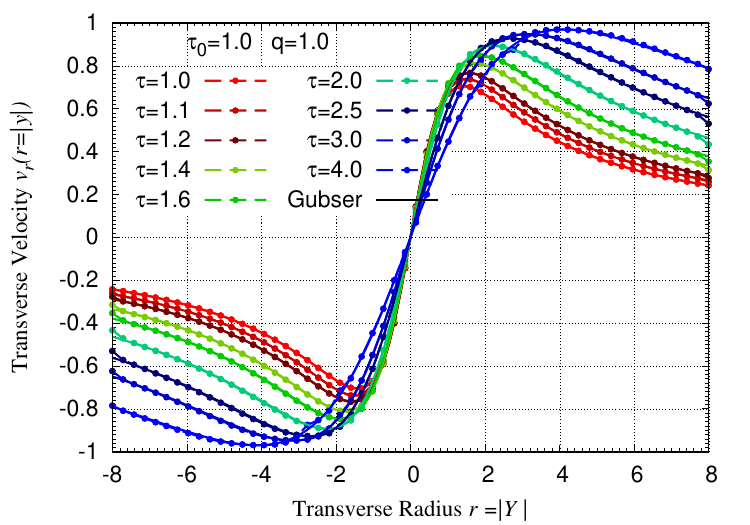}
    \caption{Our hydrodynamic simulation of an ideal fluid with the numerical implementation presented in the previous section, compared to the analytical result obtained from Gubser flow. The dashed-dotted lines are from our simulation while the solid lines are analytical results. The upper (lower) panel shows the comparison with transverse expansion parameter of $q=0.2$ ($q=1.0$).}
    \label{eq:Hydro_Gubser}
\end{figure}

In order to perform a comparison to an analytical solution, we start with the initial hydrodynamic condition $e_{\rm Gubser}(\tau_0,r)$ and $v_{r,\rm Gubser}(\tau_0,r)$ with $\tau_0=1$. The transverse expansion parameter is typically chosen as $q\simeq {\rm fm}^{-1}$ and we choose $q=0.2$ and $q=1.0$ for tests in the simulation.
We present our comparison to the analytical solution in Fig.~\ref{eq:Hydro_Gubser}. This comparison manifests our hydrodynamic code as a working code.

\pagebreak
\section{More results for a non-flowing medium}~\label{app:extra_static}
\begin{figure}[h!]
    \centering
    \includegraphics[width=0.495\textwidth]{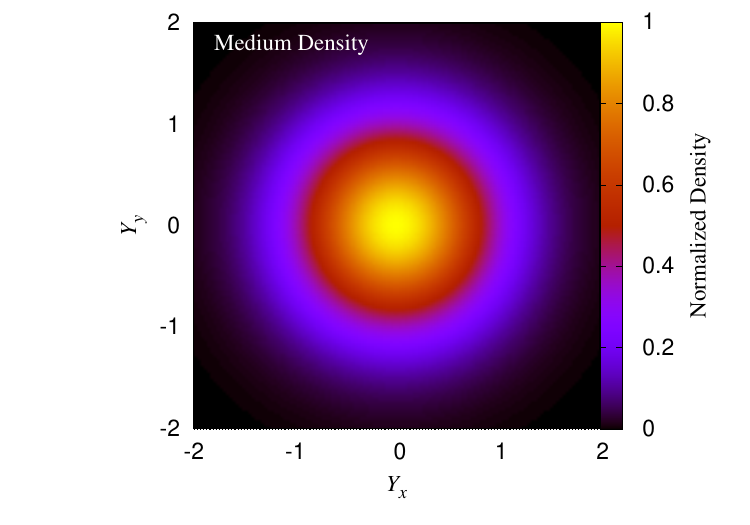}
    \includegraphics[width=0.495\textwidth]{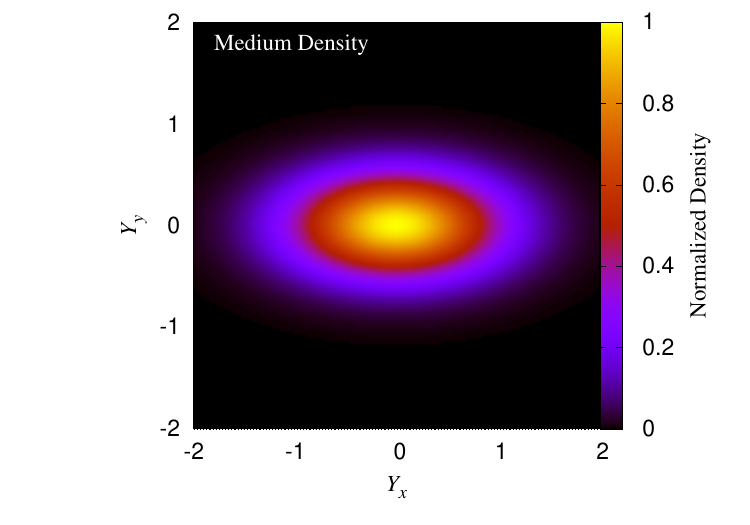}
    \caption{Normalized number density $\rho(\Y)/\rho(0)$ of an isotropic medium (left, default in calculations, if not stated otherwise) with $\mu_x=1$, $\mu_y=1$ and an anisotropic medium (right) with $\mu_x=1$, $\mu_y=2$.}
    \label{fig:medium_pic}
\end{figure}

In this Appendix, we provide some further results for the evolution of the hard probe in a non-flowing/static medium. 
First, we provide the heatmap representation of the density for the case that is studied in the main text in Fig.~\ref{fig:medium_pic} (left). In this Appendix, we repeat the numerical exercise for the case of an anisotropic medium profile, given in Fig.~\ref{fig:medium_pic} (right).

\begin{figure}[h!]
    \centering
    \includegraphics[width=0.49\textwidth]{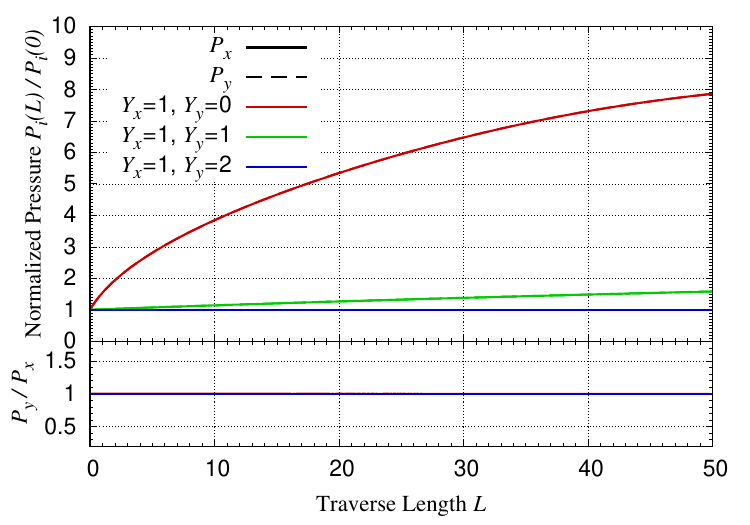}
    \includegraphics[width=0.49\textwidth]{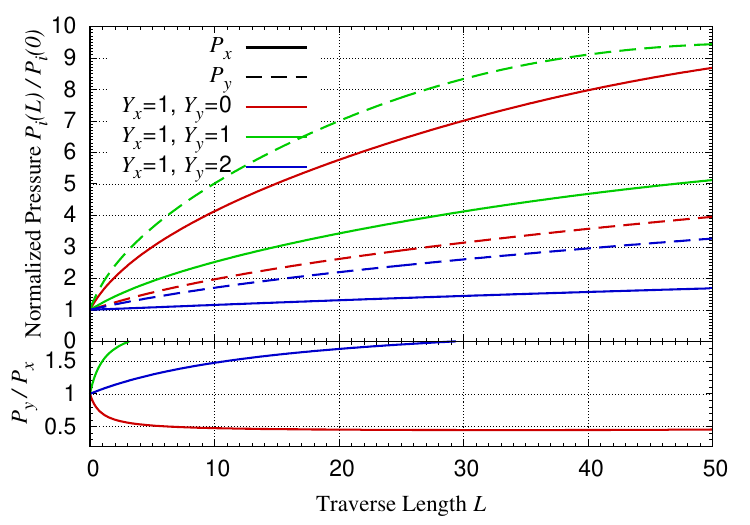}
    \caption{Pressure evolution for the anisotropic medium $\mu_x=1, \mu_y=2$, see Fig.~\ref{fig:medium_pic}, equivalent to the case studied in Fig.~\ref{fig:pressure_Y=X}.}
    \label{fig:pressure_Y_ne_X}
\end{figure}

In Fig.~\ref{fig:pressure_Y_ne_X} we show equivalent plots to the ones in Fig.~\ref{fig:pressure_Y=X}. Again, without accounting for the full dependence on the gradients (left) the pressure is the same in both directions, while the full master equation (right) leads to an anisotropy, consistent with the main text's result.

\begin{figure}[h!]
    \centering
    \includegraphics[width=0.86\textwidth]{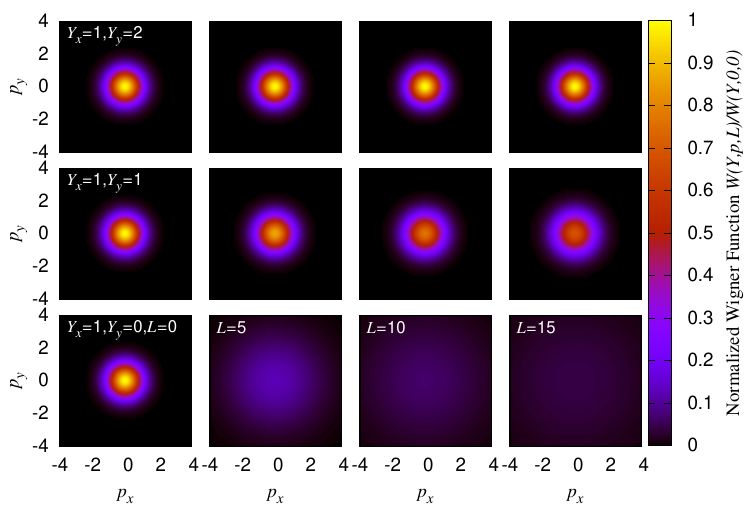}
    \includegraphics[width=0.86\textwidth]{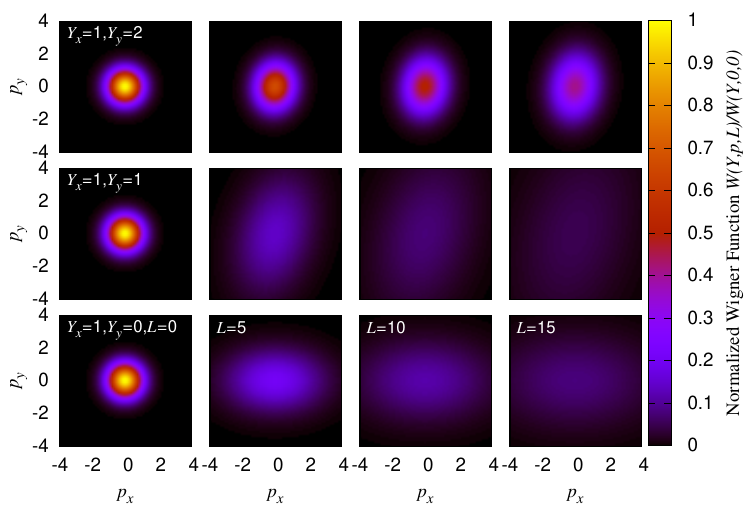}
    \caption{Same results as in Fig.~\ref{fig:strobo_RHS_1}, but now for anisotropic medium $\mu_x=1$, $\mu_y=2$. The evolution is without (upper panel) and with (lower panel) the gradient correction to the diffusion.}
    \label{fig:strob_app}
\end{figure}

In Fig.~\ref{fig:strob_app}, we provide the stroboscopic evolution of the momentum distribution for the master equation accounting only for the gradient dependence in $\hat q$ (upper) and the evolution with the full master equation (lower), similar to Fig~\ref{fig:strobo_RHS_1} but with an anisotropic medium with density parameters $\mu_x=2$, $\mu_y=1$.  Comparing to the results in the main text, one comes to the same conclusion, i.e. that a full dependence on the gradients is necessary to describe the azimuthal structure of jets.

\begin{figure}[h!]
    \centering
    \includegraphics[width=0.7\textwidth]{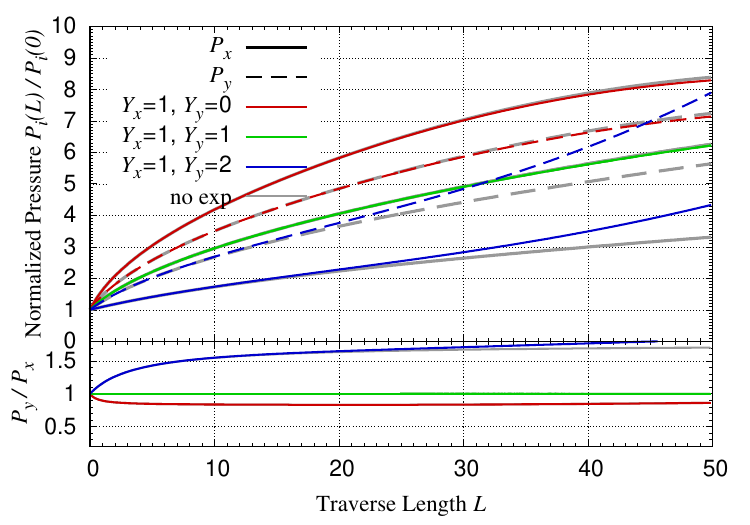}
    \caption{Equivalent plot to the one in Fig.~\ref{fig:pressure_Y=X}, but for an expanding system. Gray curves are benchmark to results we obtained without expansion, the same as in Fig.~\ref{fig:pressure_Y=X}.}
    \label{fig:appB_1}
\end{figure}

Besides the discussions on anisotropic medium above, it is also interesting to look at the evolution with expansion.
In Fig.~\ref{fig:appB_1}, we repeat the pressure plot shown in the main text but now allow for the medium to expand such that the expansion term $\p\cdot\tvec{\na}_{\Y}/E \ne 0$ in Eq.~(\ref{eq:numerics_dif_eq}). Here the major difference with respect to Fig.~\ref{fig:pressure_Y=X} is the non-linear late-time behavior of the blue curve, which is purely induced by the finite size of the lattice, and can be improved with a larger lattice size and resolution of simulations in position space $\Y$. 
The results without expansion are plotted as gray color in the figure.
Ignoring this feature, one can see that even in an expanding system the behavior is qualitatively the same as seen in the static case, due to the large energy of the jet $E_{jet}\gg 1$ and suppression of the transverse velocity $\pv/E\ll 1$, thus the expansion term $\pv\cdot\tvec{\na}_{\Y}/E \ll 1$.

\end{document}